\documentclass[fleqn,usenatbib]{mnras}

\usepackage[T1]{fontenc}
\usepackage{ae,aecompl}

\usepackage{graphicx}	
\usepackage{amsmath}	
\usepackage{amssymb}	
\usepackage{mwe}
\usepackage{todonotes}
\usepackage{xcolor}
\usepackage{hyperref}

\usepackage{enumitem}
\usepackage{lscape}
          \usepackage{pdflscape}
          \usepackage{morefloats}
			\usepackage{booktabs} 
              \usepackage{multirow}
              \usepackage{array}
              \usepackage{longtable}
               \usepackage{tabu}
                \usepackage{threeparttable} 
                 \usepackage{threeparttablex} 
                 \usepackage{rotating}
                  \usepackage{xtab}
                   \usepackage{multicol}
                   \usepackage{adjustbox}
                    
\usepackage{upgreek}
\usepackage{units}
\usepackage{caption}
\usepackage{subcaption} 

\newcommand{\kms}{${\rm km~s}^{-1}$}

\usepackage{newtxtext,newtxmath}

\DeclareRobustCommand{\ion}[2]{%
\relax\ifmmode
\ifx\testbx\f@series
{\mathbf{#1\,\mathsc{#2}}}\else
{\mathrm{#1\,\mathsc{#2}}}\fi
\else\textup{#1\,{\mdseries\textsc{#2}}}%
\fi}

\title {The ALMaQUEST Survey XIV: do radial molecular gas flows affect the star-forming ability of barred galaxies?}

\author[L. M. Hogarth et al.]{\Large \parbox{\textwidth}{
L. M. Hogarth$^{1}$\thanks{E-mail: l.hogarth.18@ucl.ac.uk},
A. Saintonge$^{1}$\thanks{E-mail: a.saintonge@ucl.ac.uk},
T. A. Davis$^{2}$,
S. L. Ellison$^{3}$,
L. Lin$^{4}$,
C. L\'opez-Cob\'a$^{4}$,
H.-A. Pan$^{5}$
and M. D. Thorp$^{6}$
}
\\
\\
$^{1}$University College London, Department of Physics \&\ Astronomy, Gower Street, London, WC1E 6BT, UK \\
$^{2}$Cardiff Hub for Astrophysics Research \&\ Technology, School of Physics \&\ Astronomy, Cardiff University, Queens Buildings, Cardiff, CF24 3AA, UK \\
$^{3}$Dept. of Physics \&\ Astronomy, The University of Victoria, Victoria, BC V8P 5C2, Canada \\
$^{4}$Institute of Astronomy \&\ Astrophysics, Academia Sinica, No. 1, Section 4, Roosevelt Road, Taipei 10617, Taiwan \\
$^{5}$Department of Physics, Tamkang University, No.151, Yingzhuan Road, Tamsui District, New Taipei City 251301, Taiwan \\
$^{6}$Argelander-Institut für Astronomie, Universität Bonn, Auf dem Hügel 71, 53121 Bonn, Germany
}
\date{Accepted XXX. Received YYY; in original form ZZZ}

\pubyear{2024}

\begin{document}
\label{firstpage}
\pagerange{\pageref{firstpage}--\pageref{lastpage}}
\maketitle

\begin{abstract}

 We investigate whether barred galaxies are statistically more likely to harbour radial molecular gas flows and what effect those flows have on their global properties. Using 46 galaxies from the ALMA-MaNGA QUEnching and STar formation (ALMaQUEST) survey, we identify galaxies hosting optical bars using a combination of the morphological classifications in Galaxy Zoo 2 and HyperLEDA. In order to detect radial molecular gas flows, we employ full 3D kinematic modelling of the ALMaQUEST $^{12}$CO(1-0) datacubes. By combining our bar classifications with our radial bar-driven flow detections, we find that galaxies classed as barred are statistically more likely to host large-scale radial gas motions compared to their un-barred and edge-on galaxy counterparts. Moreover, the majority of barred galaxies require multi-component surface brightness profiles in their best-fit models, indicative of the presence of resonance systems. We find that galaxies classed as barred with radial bar-driven flows (``barred + radial flow'' subset) have significantly suppressed global star-formation efficiencies compared to barred galaxies without radial bar-driven flows and galaxies in the other morphological sub-samples. Our ``barred + radial flow'' subset galaxies also possess consistently centrally concentrated molecular gas distributions, with no indication of depleted gas mass fractions, suggesting that gas exhaustion is not the cause of their suppressed star formation. Furthermore, these objects have higher median gas mass surface densities in their central 1~kpc, implying that a central gas enhancements do not fuel central starbursts in these objects. We propose that dynamical effects, such as shear caused by large-scale inflows of gas, act to gravitationally stabilise the inner gas reservoirs. 

\end{abstract}

\begin{keywords}
galaxies: kinematics and dynamics -- galaxies: structure -- physical data and processes: molecular data 
\end{keywords}



\section{Introduction}
\label{sec: intro}

Galactic bars are increasingly believed to be a critical stage in the secular evolution of disc galaxies in the present-day Universe \citep{masters11,chown19,geron21,geron23}. They provide an effective mechanism for the mass migration of molecular gas, which is driven inwards by the axi-asymmetric structure. This process is well-understood and detailed in studies like \citet{combes91}, \citet{bertin14}, \citet{sormani15} and \citet{krumholtz15}, which describe how bars exert powerful torques in host galaxies that drive gas onto resonant orbits with epi-cyclic frequencies commensurate with the pattern frequencies of the bars. This generally causes gas in outer-discs to fall inwards and results in the formation of resonant ring structures in the host galaxy, with the strongest resonances being at co-rotation and at the Lindblad resonances \citep{shlosman89, combes01}. Multiple direct observations of resonant ring structures using interferometry with high spatial resolutions support these models \citep[e.g.][]{olsson10, davis18, topal16, lu22}. However, observational studies with measurements directly capturing the radial motions of gas, as opposed to measurements of its spatial distribution, are more sparse. This is largely due to the resolution and sensitivity required to accurately separate non-circular gas motions from the rotation of the gas disc. With data that meet these criteria, the presence of non-circular kinematics is often estimated by assuming velocity components are Gaussian, and fitting asymmetric profiles as a superposition of Gaussians \citep[e.g.][]{lu22}, but can also be inferred more comprehensively by full three-dimensional (3D) kinematic modelling of the molecular gas \citep[e.g.][]{lelli22}.

\par In reaction to the forcing frequencies of bars, simulations in the literature anticipate central starburst activity combined with quenching of the disc, resulting from the bulk inflow of molecular gas \citep[e.g.][]{coelho11, spinoso17}. Generally these simulations follow a  ``compaction scenario'' \citep{tacchella16}, whereby galaxies cycle through active and passive star-forming phases, regulated by inflows of molecular gas and subsequent outflows once intense star formation is triggered. \citet{tacchella16} finds that galaxies at z $\sim$ 2 will oscillate along the main sequence on timescales $\rm \approx 0.4~t_{H}$ (where $\rm t_H$ is the Hubble Time). This scenario has also been simulated in the context of the Central Molecular Zone (CMZ) of a Milky Way-like galaxy by \citet{krumholtz15}. Their model predicts that the CMZ will cycle through phases above and below the star-forming main sequence on a timescale of $\rm \sim 17.5~Myr$ at resolutions of $\rm \approx 100~pc$. The short timescales of star formation and quiescence in this model are also in agreement with the observational study conducted by \citet{ellison11}, who measure elevated chemical abundances in the centres of low-mass barred galaxies. This indicates that these galaxies have experienced some past central enhancement in their star-formation rates, which was short-lived compared to the lifetime of their bars.

However, there is some contention between observational studies that have found that bars can both suppress and enhance the star-formation efficiencies (SFEs) of host galaxies. While multiple studies have confirmed elevated central molecular gas concentrations in barred galaxies \citep[driven inwards by their forcing frequency at a rate faster than the gas is consumed e.g.][]{sakamoto99,jogee05,kuno07,yu22}, studies quantifying SFE are significantly more varied. \citet{saintonge12}, for example, find no significant difference in the global SFEs of barred galaxies compared to a sample of un-barred galaxies. They acknowledge, however, that they may be averaging over galaxies undergoing different kinematic processes. This is reinforced by \citet{jogee05}, who suggest that SFEs of barred galaxies vary depending on the stage of their bar-driven gas inflow; that only galaxies hosting later-stage flows, once most molecular gas has been driven into the circumnuclear region, have enhanced SFEs. Studies such as \citet{heitsch06}, \citet{kruijssen14}, \citet{davis14} and \citet{meidt20}, however, have shown that strong non-circular motions stabilise molecular clouds against collapse or disperse them entirely, especially in the dense circumnuclear regions of galaxies. In the context of bars, suppression of SFEs have been directly observed \citep[e.g.][]{fumi18, fumiya20, fumiya23}, but conversely, star-formation enhancement in bars and circumnuclear regions have also been reported \citep[e.g.][]{watanbe11, wang12, salak17}, potentially reflecting the need to sub-divide barred galaxies based on their kinematic phase as suggested by \citet{jogee05} and \citet{saintonge12}. 

\par In this study, we search for evidence of the effects bars have on the motion, distribution, and star-formation efficiency of molecular gas across a sample of nearby galaxies. More specifically, we seek to answer the following two questions:

\begin{itemize}[leftmargin=15pt, itemsep=5pt]

    \item[1.] Can we find a statistical relationship between the presence of an optical bar and the detection of radial molecular gas motions?

    \item[2.] How does the detection of non-circular molecular gas motions influence the SFEs of barred galaxies and their global properties?

\end{itemize}

\par By attempting to answer these questions, we aim to devise a method for identifying radial molecular gas flows in a sample of galaxies and to compare those radial flow detections to global galaxy properties (e.g. an optical bar). Furthermore, we want to address the conflicting conclusions in the literature as to whether a bar suppresses or enhances SFE in its host galaxy. We will address this by studying how ongoing radial gas motions impacts star-formation activity, and how that relates to a galaxy's position on the star-forming main sequence.

Throughout this paper we adopt a standard $\Lambda$CDM cosmology with $\rm H_0=70\ $\kms~Mpc$^{-1}$, $\rm \Omega _{m_0}=0.3$, $\rm \Omega _{\Lambda}=0.7$ and a \citet{chabrier03} IMF. 


\section{Sample selection \& data}
\label{sec: selectionanddata}

\subsection{ALMaQUEST} 
\label{subsec: ALMaQUEST}

\begin{figure}
    \centering
    \includegraphics[width=\linewidth]{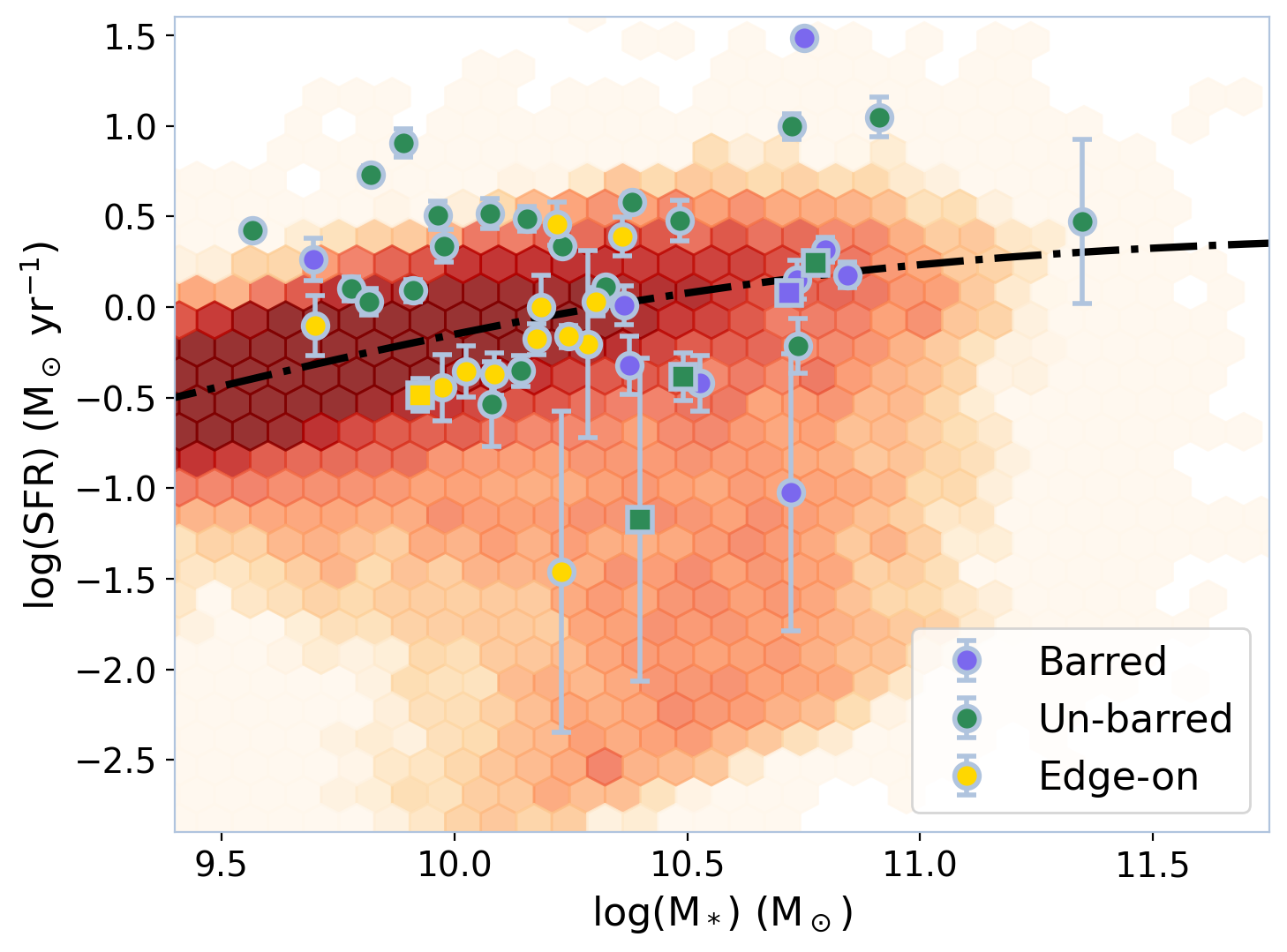}
 \caption{SFR-$\rm M_*$ plot illustrating the position of the ALMaQUEST sample with respect to the star-forming main sequence. SFRs are taken from the GSWLC-X2 catalogue \citep{salim16, salim18} where available and stellar masses are taken from the NSA catalogue \citep{blanton_roweis07, blanton11}. The five objects not covered by GSWLC-X2 (square markers) are supplemented with SFRs from WISE \citep{wright10}. The GSWLC-X2 catalogue at $0.01 \leq z \leq 0.05$ is also illustrated in its entirety as a heat map and the star-forming main sequence \citep[as derived by][]{saintonge22} is indicated as a black dashed line. We colour the markers by the morphological class of the galaxy (``barred'', ``un-barred'' or ``edge-on''). Our classification procedure is detailed in Section~\ref{subsec: morph class}.}
     \label{fig: ms_plot}
\end{figure}

The ALMA-MaNGA QUEnching and STar formation (ALMaQUEST) sample \citep{lin20,ellison24} consists of 47 galaxies selected from DR14 and DR15 of the Mapping Nearby Galaxies at Apache Point Observatory survey \citep[MaNGA;][]{bundy15}, which span a wide range of specific star-formation rates and have stellar masses ($\rm M_*$) in the range $\rm 10 \lessapprox \log (M_*/M_\odot ) \lessapprox 11.5$ (including the green valley, main sequence and starburst regimes; see Figure~\ref{fig: ms_plot}). The Atacama Large Millimeter Array (ALMA) $^{12}$CO(1-0) (rest frequency 115.271204 GHz) observations were collected from four individual ALMA programmes: 2015.1.01225.S, 2017.1.01093.S, 2018.1.00558.S (PI: Lin), and 2018.1.00541.S (PI: Ellison). We note that 1 object in the sample has no CO(1-0) detection.

\par All ALMA observations were taken in the C43-2 configuration (with a synthesised beam of 2.5\arcsec) to be comparable to the angular resolution of the optical integral-field spectroscopic data of the MaNGA survey. Typically, ALMaQUEST data is spatially resolved on physical scales $\approx$1.5~kpc (ranging between 0.9 to 6~kpc across the sample), with the largest structure the data are sensitive to being $\approx$14~kpc. The spectral set-up includes a high-resolution spectral window with a channel width of $\approx$10~\kms\ targeting $\rm ^{12}$CO(1-0). The integration time per object ranges from 0.2 to 2.5 hours to ensure a signal-to-noise ratio greater than 3 for more than 50\% of the spatial pixels (spaxels) where the MaNGA H$\upalpha$ signal-to-noise ratio is also greater than 3. 

\par In order to generate moment maps and calculate total H2 masses, we mask the $^{12}$CO(1-0) datacubes by smoothing the original datacubes with Gaussian kernels with widths of $\rm 1.5 \times {Bmaj}_{pix}$ spaxels spatially (where $\rm {Bmaj}_{pix}$ is the width of the ALMA synthesised beam's major-axis in pixels) and 4 channels spectrally. Values lower than the standard deviation of the original datacube are set to 0 and all those above are set to 1. The original datacubes multiplied by these masks are what we use to calculate the total CO(1-0) luminosities ($\rm L_{CO}$), with the relation:

\begin{equation}
    \rm L_{CO} = 3.25 \times 10^7 \times S_{CO} \Delta v \times {\upnu_{sys}}^{-2}\times {D_{L}}^2\times {(1 + z)}^{-3}\ ,
\end{equation}

\hspace{2pt} where $\rm S_{CO}$ is the total CO intensity obtained by summing over the masked cube, $\rm \Delta v$ is the channel width, $\rm \upnu_{sys}$ is the redshifted frequency of the CO(1-0) line, $\rm D_{L}$ is the luminosity distance and $\rm z$ is the redshift of the object (obtained from the NASA-Sloan Atlas catalogue; NSA). The total molecular gas mass of each object is then calculated using the constant Galactic conversion factor $\rm \upalpha _{CO} = 4.3~M_\odot\ {(K~km~s^{-1}~pc^2)}^{-1}$, which includes a correction for the presence of heavier elements/molecules as detailed in Section~\ref{sec: intro}. All galaxies in the sample have near-solar metallicities (calculated with the $\rm 12 + \log\left( O/H \right)$ proxy), with a median gas-phase metallicity of $8.69 \pm 0.05$, where we use the measurements of \citet{lin20} inferred with the O3N2 calibrator derived by \citet{pettini04}.

\subsection{Star-formation rates \& stellar masses}
\label{subsec: sfr}

\subsubsection{MaNGA data products}
\label{subsec: manga}

As referred to in Section~\ref{subsec: ALMaQUEST}, the MaNGA survey \citep{bundy15} is composed of integral-field spectroscopic (IFS) observations of galaxies in the local Universe ($\rm z\lessapprox0.15$) with coverage from 3600 to 10400~\AA. We use optical emission line fluxes from the MaNGA Pipe3D pipeline derived from Data Release 17 \citep{sanchez16, sanchez18, sanchez22}. These are corrected for internal extinction by adopting an intrinsic H$\upalpha$/H$\upbeta$ = 2.86 and a Milky Way extinction curve \citep{cardelli89}. We follow the masking procedure detailed in \citet{ellison21}, which makes restrictions on the optical spaxels used in their study combining MaNGA and ALMaQUEST data. This process can be summarised as follows:

\begin{itemize}

    \item[1.] Spaxels are required to have $\rm S/N>2$ in H$\upalpha$, H$\upbeta$, [OIII]$\uplambda$5007 and [NII]$\uplambda$6584 maps.
    \item[2.] Each spaxel must fall in the star-forming portion of the Baldwin, Phillips and Terlevich (BPT) diagram \citep{baldwin81} using the \citet{kauffmann03} classification.
    \item[3.] Spaxels require a H$\upalpha$ equivalent width (EW) > 6~\AA.
    \item[4.] Using stellar-mass surface density maps ($\rm \Sigma _{M_*}$) from the Pipe3D pipeline, each spaxel must have $\log \rm \Sigma _{M_*} > 7$ (masking anomalously small values).
    
\end{itemize}

This procedure effectively masks low S/N spaxels and those contaminated by active galactic nuclei (AGN). We calculate star-formation rates (SFRs) from the extinction corrected, masked H$\upalpha$ maps using the relation given in \citet{kennicutt98}, so that:

\begin{equation}
    \rm SFR~[M_\odot~{yr}^{-1}] = 7.9 \times {10}^{-42}~L (H \upalpha)~[ergs~s^{-1}]\ ,
\end{equation}

\hspace{2pt}  where $\rm L (H \upalpha)$ is the luminosity of H$\upalpha$ emission. We note, however, that the masking process used to derive these SFR maps does significantly mask the central regions of some objects in the ALMaQUEST sample. In order to supplement MaNGA IFS, therefore, we also utilise independent global star-formation rates (detailed in Sections~\ref{subsubsec: gswlc} and \ref{subsubsec: wise}).

\begin{figure*}
    \centering
    \includegraphics[width=\linewidth]{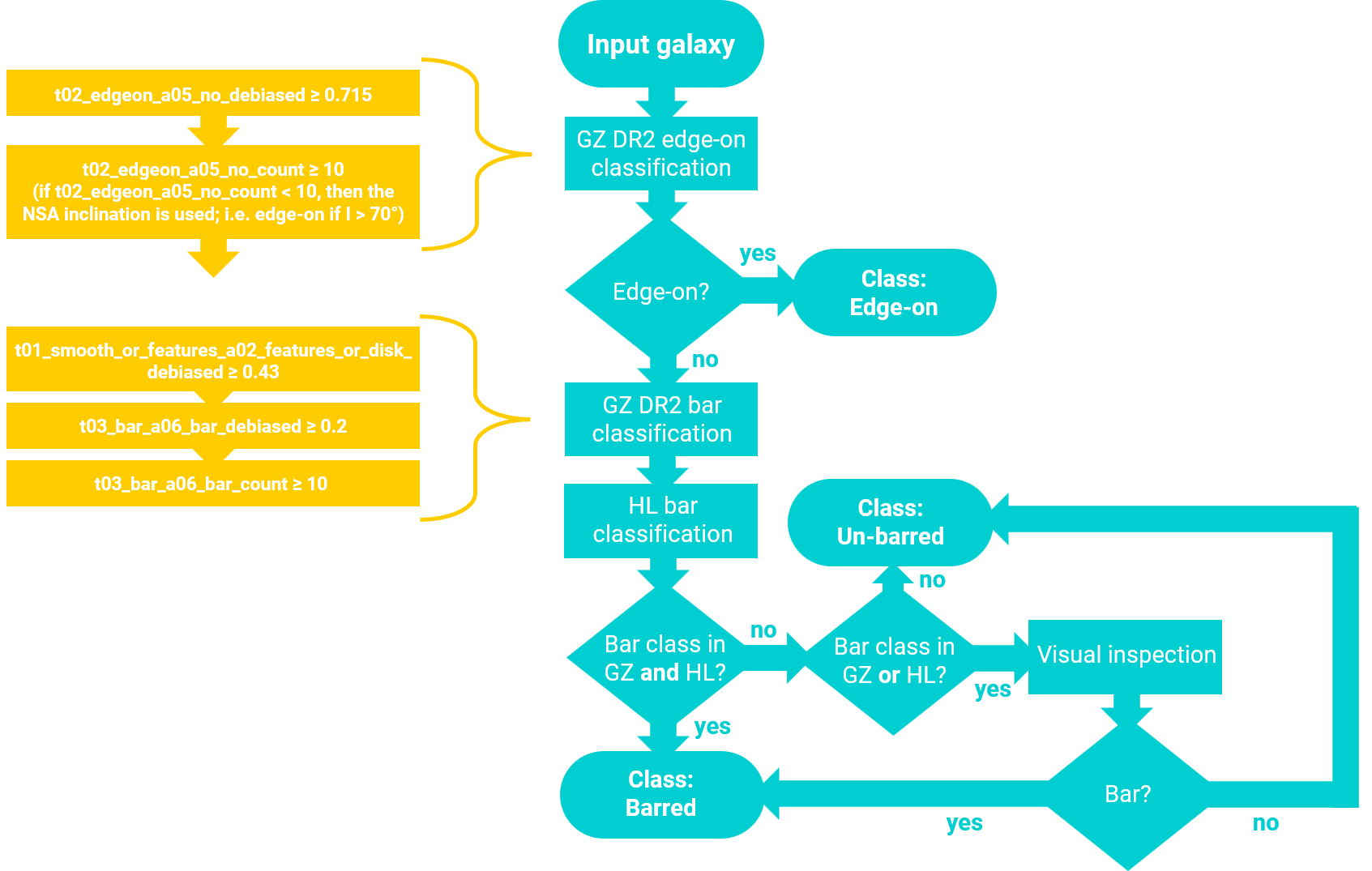}
     \caption{Flowchart illustrating our bar classification process using both GZ2 \citep{willett13} and HL \citep{makarov14}.}
     \label{fig: bar_class}
\end{figure*}

\subsubsection{GSWLC}
\label{subsubsec: gswlc}

The GALEX-SDSS-WISE Legacy Catalogue \citep[GSWLC;][]{salim16, salim18} contains measured properties of over 700,000 galaxies within the GALEX footprint and with Sloan Digital Sky Survey (SDSS) redshifts between 0.01 and 0.3. The catalogue is split into two versions: GSWLC-1 and GSWLC-2. We use SFRs from GSWLC-X2, which uses joint UV+optical+mid-infrared (IR) spectral energy distribution (SED) fitting, using Wide-field Infrared Survey Explorer (WISE) 22-micron photometry (see Section~\ref{subsubsec: wise}).

\par GSWLC-X2 covers 41 of the 46 objects we use from ALMaQUEST, illustrated in Figure~\ref{fig: ms_plot} alongside the GSWLC-X2 catalogue at $0.01 \leq z \leq 0.05$.

\subsubsection{WISE}
\label{subsubsec: wise}

For the remaining 5 objects from ALMaQUEST that are not included in the GSWLC-X2 catalogue, we use SFRs derived only from the WISE 22-micron photometry in the AllWISE database \citep{wright10, cutri14}. We employ the method outlined in \citet{janowiecki17} to calculate SFRs from W4-band fluxes ($\rm {SFR}_{W4}$), using SDSS redshifts to calculate the luminosities of the W1 and W4 bands and using the calibration of \citet{jarrett13} with a correction for stellar mid-infrared (MIR) contamination based on the W1-band luminosities:

\begin{equation}
    \rm {SFR}_{W4}~[M_\odot~y^{-1}] = 7.5 \times 10^{-10} \times (L_{W4} - 0.044~L_{W1})~[L_\odot]\ ,
\end{equation}

\hspace{2pt} where $\rm L_{W1}$ and $\rm L_{W4}$ are the W1 and W4-band luminosities, respectively. We show the SFRs derived from WISE in Figure~\ref{fig: ms_plot} alongside those from the GSWLC-X2 catalogue.

\subsubsection{NSA}

For all of our objects, we use stellar masses calculated from K-corrected fits of Sersic fluxes from the NASA-Sloan Atlas \citep{blanton_roweis07, blanton11}. The NSA catalogue contains a wealth of parameters fit to photometry of local galaxies observed using SDSS. We use the parameter \texttt{nsa\_sersic\_mass} from the \texttt{drpall-v3\_1\_1.fits} file released with DR17 of SDSS.

\subsection{Morphological sub-samples}
\label{subsec: morph class}

For this investigation, we require a robust method for identifying barred galaxies within the sample. To achieve this, we use two catalogues that contain morphological data: Galaxy Zoo Data Release 2 \citep[GZ2;][]{willett13} and HyperLEDA \citep[comprised of the HyperCat and Lyon-Meudon Extragalactic Database/LEDA, henceforth referred to as HL;][]{makarov14}. We define three morphological sub-samples; barred, un-barred and edge-on (where edge-on galaxies are too highly inclined to establish whether a bar is present). All galaxies in our sample are included in the HL database, but 5 are missing from GZ2. The procedure we devised using GZ2 and HL is illustrated as a flowchart in Figure~\ref{fig: bar_class}.  

\par For each galaxy in the sample, we first use GZ2 to determine whether the galaxy is edge-on \citep[we use redshift debiased values from GZ2 throughout this procedure, as defined by][]{willett13}. We find using the GZ2 classifications of the edge-on sub-sample more accurate than using an inclination cut as the classification is based on whether the galaxy features are visible to the participants. Applying an arbitrary inclination cut led to some galaxies being classified as edge-on when their features were still discernible (e.g. a central bar was visible). Details of the combination of GZ2 parameters used in this classification are included in Figure~\ref{fig: bar_class}. We require a lower count thresholds in the \texttt{t02\_edgeon\_a05\_no\_count} and \texttt{t03\_bar\_a06\_bar\_count} parameters than the recommendation (i.e. we use $\geq 10$ instead of $\geq 20$), as we find low count numbers for this parameters for many objects in the sample. However, after a visual assessment, we still find these classifications reliable with the lower count threshold. If the count number is <10 for the \texttt{t02\_edgeon\_a05\_no\_count}, we instead use an inclination cut of $>70^\circ$ (where inclinations are taken from the NSA catalogue) to determine if a galaxy is edge-on. We also use the lower limit for \texttt{t03\_bar\_a06\_bar\_debiased} of $\geq 0.2$, where \citep{willett13} find $\leq 0.2$ correlates strongly with galaxies classified as un-barred by the reference sources \citet{nair10, devaucouleurs91}. Moreover, this lower limit is less biased towards strongly barred galaxies (i.e. it is more sensitive to weakly barred objects).

\begin{figure*}
    \centering
    \includegraphics[width=.9\linewidth]{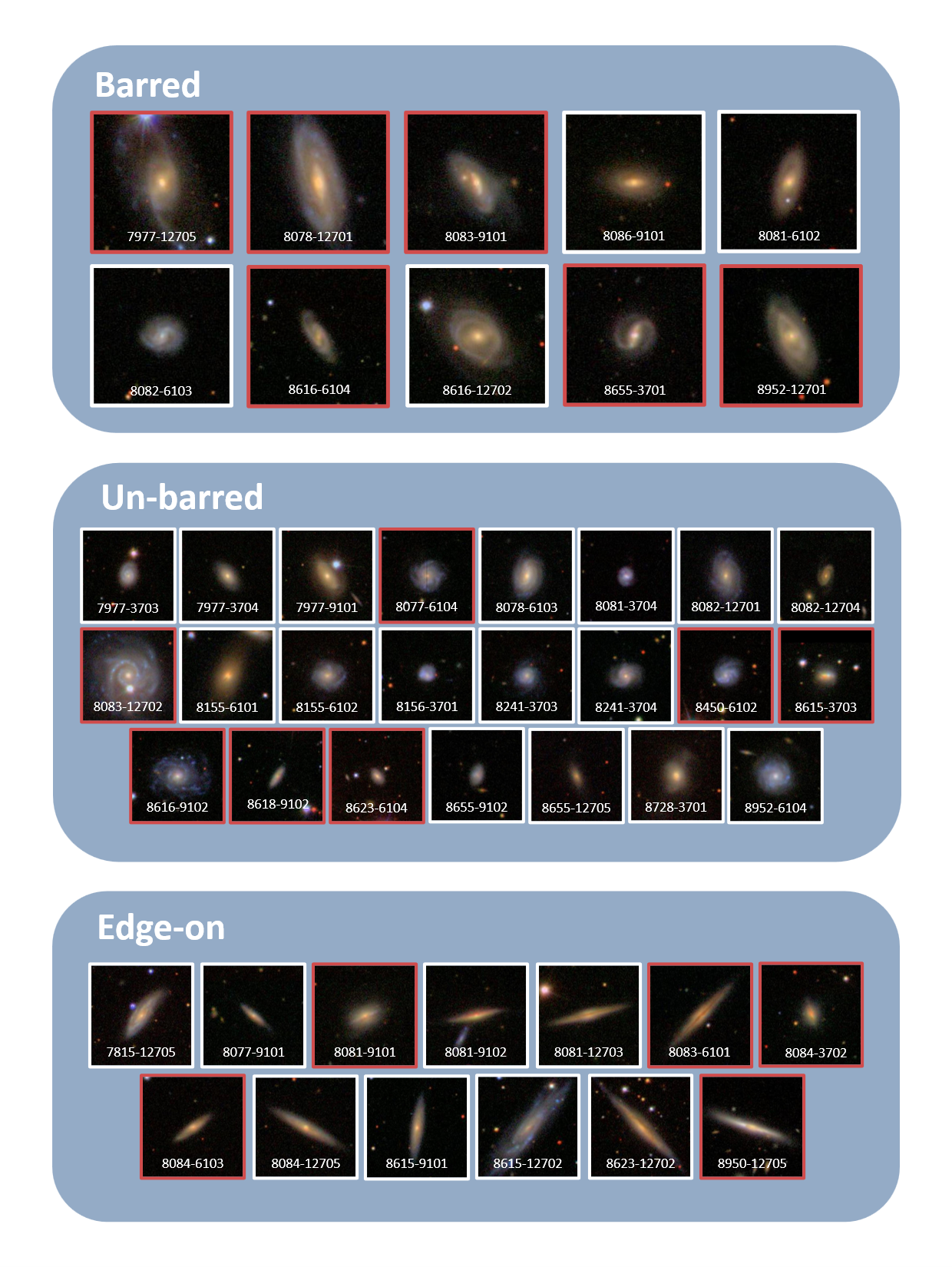}
    \caption{Optical SDSS $gri$ composite images of the 46 galaxies in ALMaQUEST, grouped by their morphological classification as detailed in Figure~\ref{fig: bar_class}. Galaxies with a red border are classed as hosting radial bar-driven flows by the process described in Figure~\ref{fig: model_class}.}
    \label{fig: optical}
\end{figure*}

\par If a galaxy is not classed as edge-on, we look at whether it is classed as having a bar in both GZ2 and HL (details of the parameters used from GZ2 are given in Figure~\ref{fig: bar_class}). If GZ2 and HL both agree that the galaxy has a bar, it is added to our barred sub-sample. Likewise, if they both agree that the galaxy is un-barred, we add it to our un-barred sub-sample. If there is a disagreement between the catalogues (or it is only present in one of the catalogues), we visually inspect the SDSS red-green-blue (RGB) image of the object and decide whether or not a bar is present. We note, however, that our classification procedure is contingent upon the visibility of features in SDSS imaging, which is dependent on both the redshift of galaxies and their size. Using redshift debiased classifications from GZ2 along with the lower-limit for bar classification should reduce this effect to some extent, but we are unable to discount it entirely given the imaging and studies available.

\begin{figure*}
    \centering
    \includegraphics[width=\linewidth]{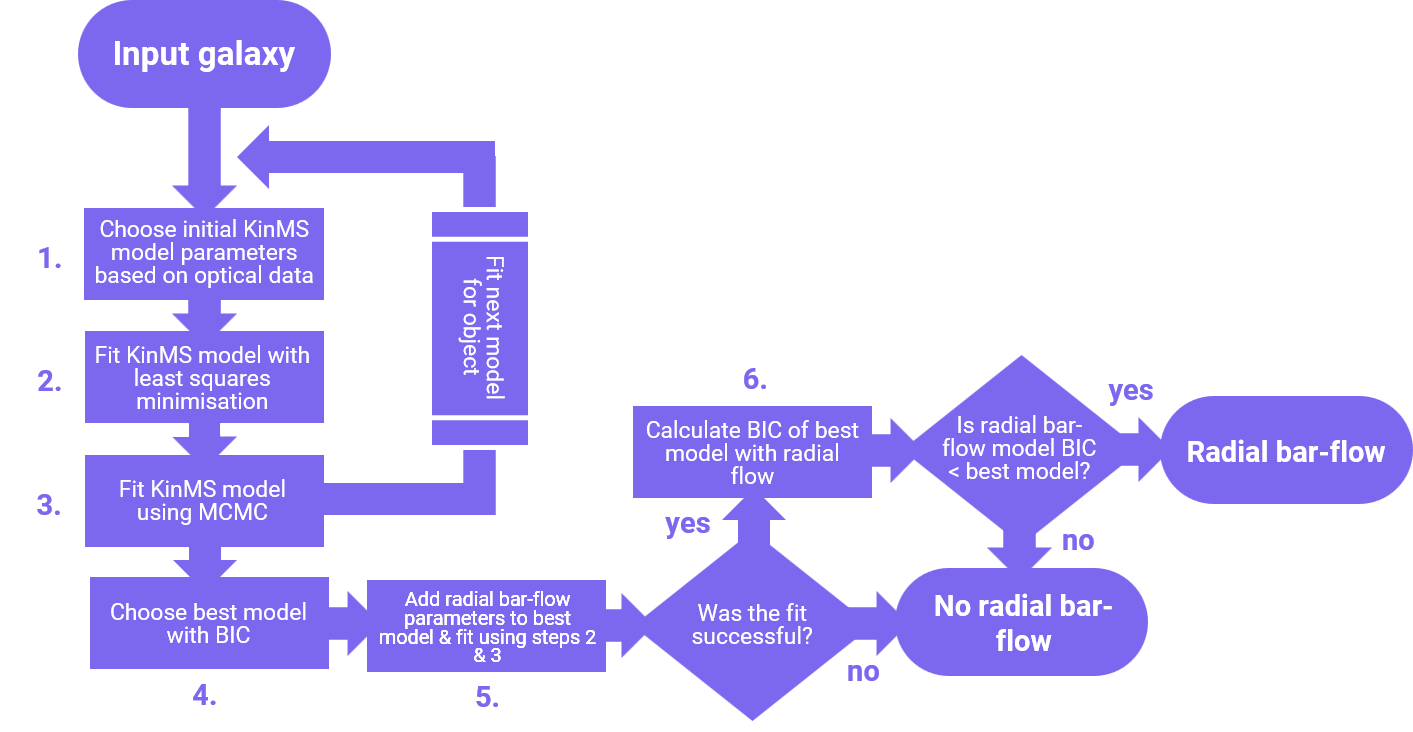}
     \caption{Flowchart illustrating our process for fitting and selecting the ``best-fit'' kinematic model for each of the galaxies in ALMaQUEST.}
     \label{fig: model_class}
\end{figure*}

\par In total, 10 objects are included in the barred sub-sample, 23 in the un-barred sub-sample and 13 in the edge-on sub-sample. These sub-samples are represented using SDSS $gri$ composite images of the galaxies in Figure~\ref{fig: optical}. From this analysis, we calculate a bar fraction of $0.30 \pm 0.08$ (where the uncertainty is the binomial uncertainty), while the bar fraction calculated for the whole GZ2 sample in \citet{willett13} is 0.35 (given that we add the requirement of bar-detection in HL, a lower bar fraction is expected). Furthermore, the ALMaQUEST sample also covers a different range of global properties ($\rm M_*$, SFR, size, etc.) compared to the full GZ2 sample. Our sub-samples are referred to as the ``morphological sub-samples'' throughout the rest of this text.


\section{Results}
\label{sec: results}


\subsection{Kinematic models} 
\label{subsec: models}

In order to create three-dimensional kinematic models of the sample, we use the process described in \citet{hogarth23}, which makes use of the KINematic Molecular Simulation tool \citep[\textsc{KinMS}; see][]{davis13, davis20, lelli22}\footnote{\url{https://github.com/TimothyADavis/KinMSpy}}\footnote{\url{https://kinms.space}}. \textsc{KinMS} allows us to create simple simulated interferometric datacubes by defining arbitrary surface brightness and velocity profiles. We also use the \texttt{KinMS\_fitter}\footnote{\url{https://github.com/TimothyADavis/KinMS_fit}} wrapper of \textsc{KinMS}, which acts as a front end for the most common fitting tasks, and provides a simple interface for defining surface brightness and velocity profiles. It also provides an interface to the \textsc{GAStimator} package\footnote{\url{https://github.com/TimothyADavis/GAStimator}}, which implements a Python Markov chain Monte Carlo (MCMC) Gibbs sampler with adaptive stepping to fit the mock interferometric datacubes generated by KinMS, with predefined surface brightness and velocity profiles, to the original datacubes.

\par For each object in ALMaQUEST, we fit six different surface brightness profiles: an exponential disc, an exponential disc~+~central hole, a Gaussian ring, two exponential discs, an exponential disc~+~a Gaussian ring and two Gaussian rings (i.e. three single-component profiles and three double-component profiles). Each exponential disc has two free parameters describing the peak surface brightness ($F_{\rm exp,\ peak}$) and scale width ($R_{\rm exp}$), while each Gaussian ring has three free parameters describing the peak surface brightness ($F_{\rm gauss,\ peak}$), the mean radius of the ring ($R_{\rm gauss}$) and the width of the ring ($\sigma _{\rm gauss}$). In order to reduce the number of parameters, in one-component profiles the peak flux is fixed to 1 and in two-component profiles the peak flux of the first component in fixed to 1, while the other is left to vary relative to the first. In addition, each model has three parameters describing the kinematic centre of the gas disc: the spatial coordinates of the dynamic centre ($x_0$, $y_0$) and the systemic velocity $V_{\rm sys}$. Two further parameters, the position angle (PA) and inclination ($i$) are used to define the orientation of the disc. The total flux ($F_{\rm gas}$) contained in the disc is also left to vary freely. In all of the models we use an arctan circular velocity curve to capture the gas rotation, which we find to be an excellent approximation of the molecular gas rotation for the majority of the objects. This profile requires two additional parameters: the maximum velocity ($V_{\rm max}$) and the turnover radius ($R_{\rm turn}$; the radius at which $V_{\rm max}$ occurs), so that the rotation velocity ($V_{\rm rot} (r)$) is described by:

\begin{equation}
    V_{\rm rot} (r) = \frac{2 V_{\rm max}}{\pi} \arctan{\left(\frac{r}{R_{\rm turn}}\right)}\ , 
\end{equation}

\hspace{2pt} where $r$ is the radius in the plane of the disc. The gas is also given a spatially-constant velocity dispersion ($\sigma _{\rm gas}$) in each model. The total number of parameters used in each model varies depending on the surface brightness model used, with 10 being the minimum and 14 the maximum.

\begin{figure*}
    \centering
    \includegraphics[width=.7\textwidth]{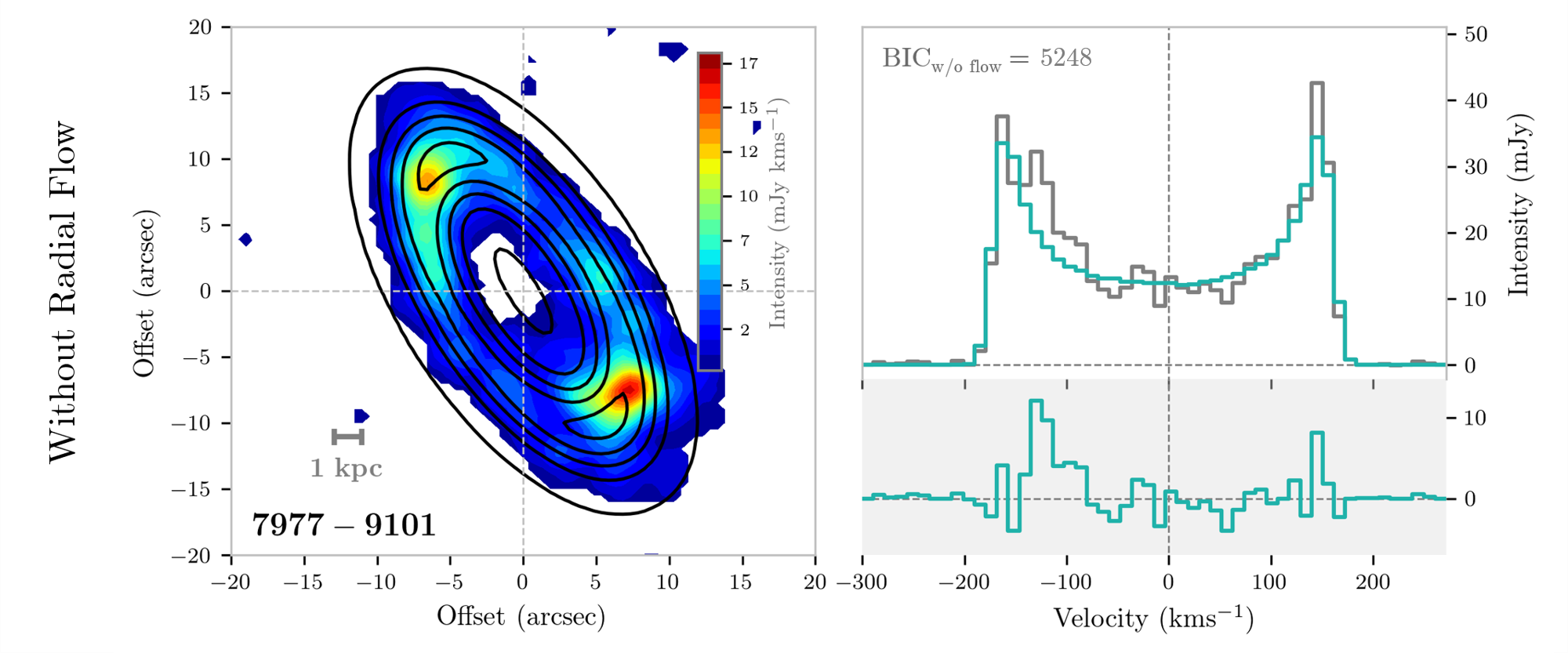}
    \caption{Example best-fit model of the ALMaQUEST galaxy with MaNGA Plate IFU 7977-9101 in our un-barred morphological sub-sample. \textbf{Left}: masked zeroth-moment map of 7977-9101 with the best-fit model overlaid in black contours. \textbf{Right}: spectrum extracted from the datacube of 7977-9101 (grey line), with the spectrum extracted from the best-fit model cube (turquoise line) overlaid. We also show the $\rm {BIC}_{w/o\ flow}$ of the model above the spectra. The lower panel shows the residuals between the data and model spectra.}
    \label{fig: 7977-9101_model}
\end{figure*}

\par In Figure~\ref{fig: model_class}, we illustrate the process used to find the best-fit model of each of the objects in the sample. Before fitting each model with MCMC, we fit an initial model using a least-squares minimisation. The MCMC is set with uniform priors with physically-motivated boundaries and run using 100,000 steps with 25 workers. Once a \textsc{KinMS} model has been fit using MCMC, we repeat steps 1-3 in Figure~\ref{fig: model_class} until all six pre-defined models have been generated and fit to the data. When all models have been fit to the input galaxy, we use the Bayesian Information Criteria (BIC) to select the best-fit model out of the six \textsc{KinMS} models. The best-fit model is the one that produces the lowest BIC, whereby more complex surface brightness models are penalised for larger parameter spaces:

\begin{equation}
    \rm BIC \equiv k \log(n)-2 \log (\widehat{L}) \ ,
\end{equation}

\hspace{2pt} where $k$ is the number of parameters of the model, $n$ is the number of spaxels in the datacube with signal and $\widehat{L}$ is the maximum value of the likelihood function defined in \textsc{GAStimator}. We reject models that have fit parameters with uncertainties (defined as the width between the $\rm 16^{th}$ and $\rm 84^{th}$ percentiles of the probability distribution function) larger than half the size of the parameter range explored (we find this works well as a method of rejecting unconverged models). We successfully find a converged best-fit model for each object in the sample.

\par The next stage in our process is to add radial bar-driven flows to the best-fit model \citep[again using the same technique outlined in][]{hogarth23}. We model radial motions induced by a central bar by using the \texttt{radial$\_$barflow} function of the \texttt{KinMS$\_$fitter} wrapper, which is based on the non-axisymmetric models described in \citet{spekkens07}. This radial bar-driven flow model simplifies the models outlined in \citet{spekkens07} by assuming that there is a bar that extends from the galaxy's centre to a radius $R_{\rm b}$, with a phase $\phi_{\rm b}$, and that the gas has constant radial and transverse velocities ($\overline{V}_{\rm r}$, $\overline{V}_{\rm t}$) within the bar's radius. $\overline{V}_{\rm r}$ represents the mean radial flow in the plane of the disc and $\overline{V}_{\rm t}$ the mean streaming speed of the gas perpendicular to $\overline{V}_{\rm r}$. The total radial flow velocity ($V_{\rm f} (r, \theta)$) is defined as:

\begin{equation} 
    V_{\rm f}(r, \theta) = 
        \begin{cases}
            V_{\rm f} (\theta), & \text{if}\ r \leq R_{\rm b} \\
            0, & \text{if}\ r > R_{\rm b} \ ,
        \end{cases}
\end{equation}

\hspace{2pt} where $\theta$ is the angle in the disc plane relative to the major-axis and $V_{\rm f} (\theta)$ is the flow velocity at radii $\leq R_{\rm b}$, which varies only with $\theta$ so that:

\begin{equation}
     V_{\rm f} (\theta) = - \sin{i} [ \overline{V}_{\rm t} \cos{\left( 2 (\theta_{\rm b}) \right)} \cos{\theta} + \overline{V}_{\rm r} \sin{\left(2 (\theta_{\rm b}) \right)} \sin{\theta} ] \ ,
\end{equation}

\hspace{2pt} where $\theta _{\rm b} = \theta - \phi _{\rm b}$ represents the angle relative to the position angle of the bar. This simplified radial bar-driven flow model is a first-order approximation of the non-circular kinematics of gas in the presence of a bar, where we assume that the bulk of the flow will happen in the circumnuclear region of the galaxy and that the flow in this region can be described with constant $\overline{V}_{\rm t}$ and $\overline{V}_{\rm r}$ (i.e. without a radial dependence). This reduces the computational time by reducing $V_{\rm f} (r, \theta)$ \textrightarrow $V_{\rm f} (\theta)$, and allows us to identify the presence of radial flows statistically over a sample of galaxies.  

\par The radial bar-driven flow approximation adds an additional four parameters to our models. At minimum, therefore, each model has 14 free parameters and at maximum 18 free parameters. For each galaxy, we re-fit our best-fit model with radial bar-flows using MCMC, using the same priors and set-up as described previously (with additional uniform priors for the radial flow parameters). If the model does not converge (based on the width of each parameter's uncertainty), we classify the object as not having a radial bar-driven flow. We also require either $\overline{V}_{\rm t}$ or $\overline{V}_{\rm r}$ to be larger than $2 \times$ the channel width (where $\rm 2 \times \Delta _{channel} \approx 22~km~s^{-1}$) as a threshold for radial flow detection. If these requirements are met, we re-calculate the BIC using the new model in step 6 (see Figure~\ref{fig: model_class}) and if $\rm{BIC}_{w/\ radial\ flow}\ < BIC_{w/o\ radial\ flow}$, the object is classed as having radial bar-driven flows (otherwise, it is classed as not having radial bar-driven flows). We note that in the majority of our barred objects classed as hosting radial bar-driven flows, we find a good alignment between the sky projection of $\phi\ _{\rm b}$ and the optical position angle of the bar \citep[echoing the results of][who find the same result using a more detailed bisymmetric flow model]{lopez-coba22}.

\begin{figure*}
    \centering
    \includegraphics[width=.6\textwidth]{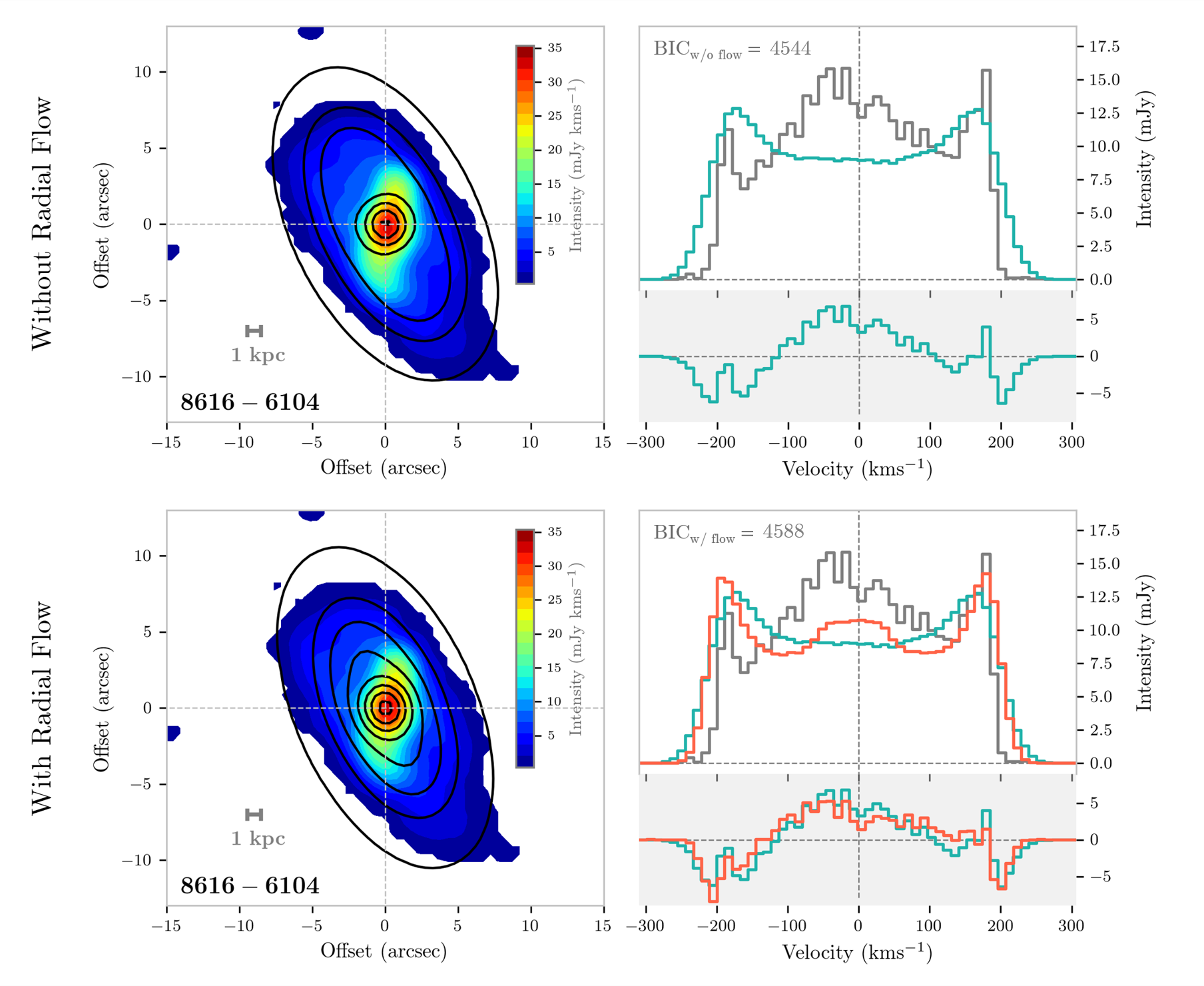}
    \caption{Example best-fit models of the ALMaQUEST galaxy with MaNGA Plate IFU 8616-6104 in our barred morphological sub-sample, without (upper panels) and with (lower panels) radial bar-driven flows. The layout of the panels is equivalent to that in Figure~\ref{fig: 7977-9101_model}, except in the panel showing the spectra of the model with radial flows. In this panel, we plot the spectrum extracted from the model without radial flows (turquoise) and with radial flows (red), to allow a visual comparison between the two. We show the BIC of both models above the spectra (i.e. $\rm {BIC}_{w/o\ flow}$ and $\rm {BIC}_{w/\ flow}$). In the case of this object, $\rm {BIC}_{w/\ flow}$ < $\rm {BIC}_{w/o\ flow}$.}
    \label{fig: 8616-6104_model}
\end{figure*}

\begin{figure*}
    \centering
    \includegraphics[width=.6\textwidth]{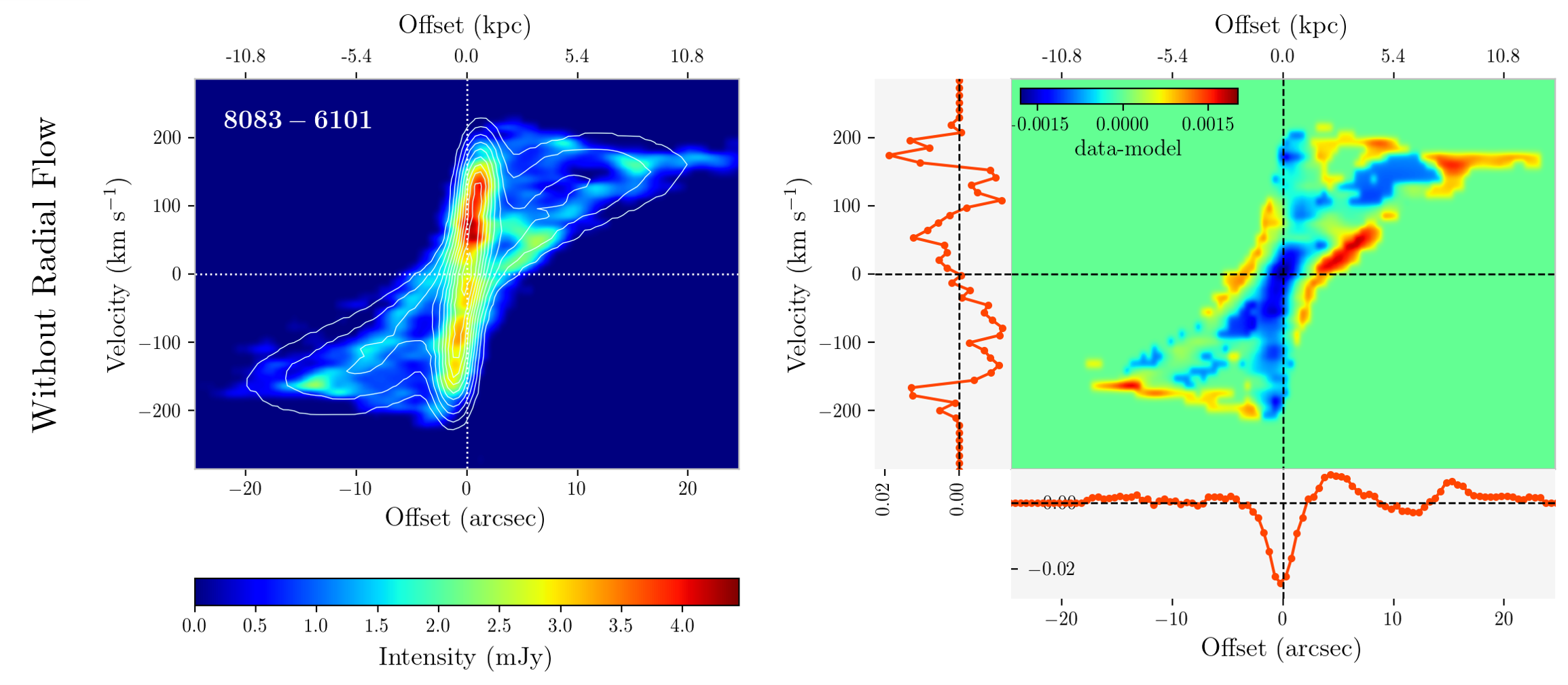}\vspace{5pt}
    \centering
    \includegraphics[width=.6\textwidth]{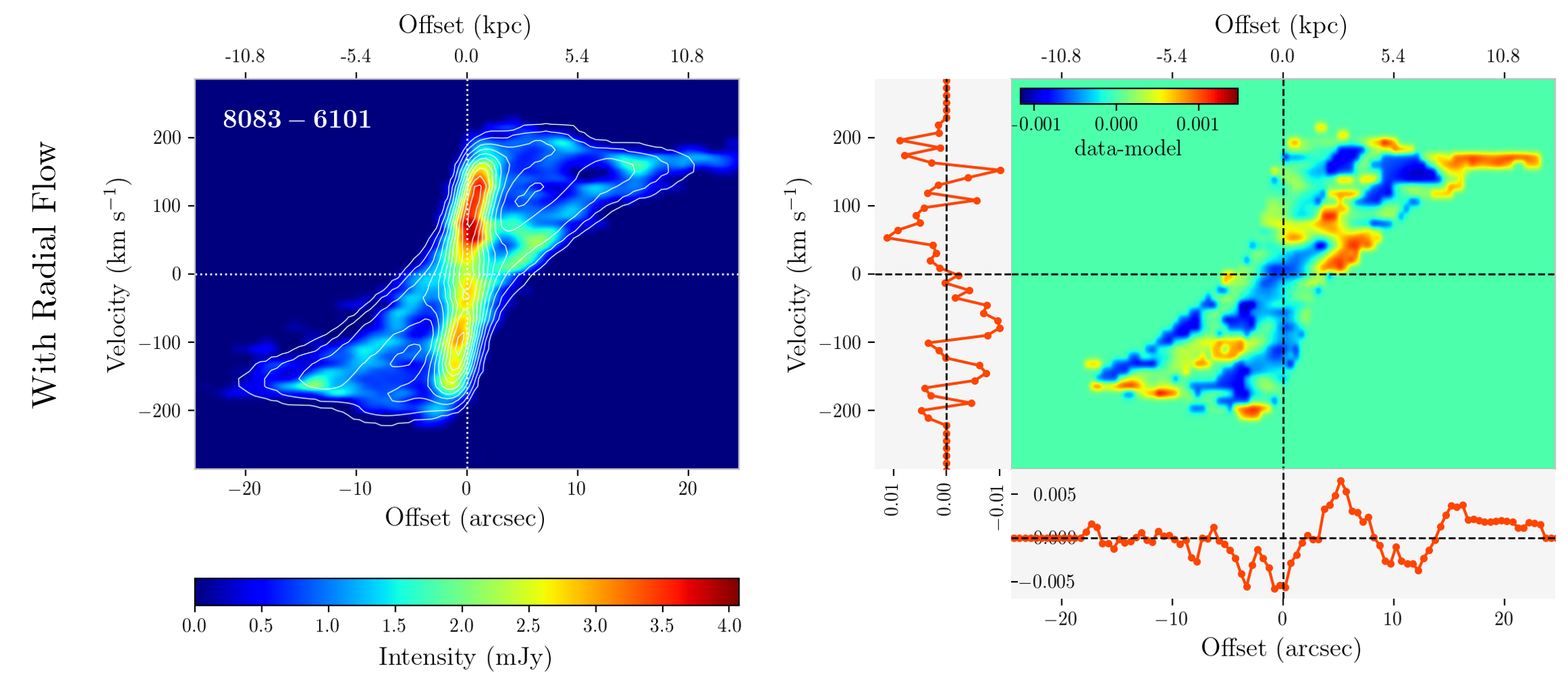}
    \caption{Example position-velocity diagrams (PVDs) illustrating the best-fit models of the ALMaQUEST galaxy with MaNGA Plate IFU 8083-6101 in our edge-on morphological sub-sample, both without (upper panels) and with (lower panels) radial flows included in the models. \textbf{Left}: PVD extracted form the original datacube of 8083-6101 (heatmap) with the PVD extracted from the model overlaid as white contours. The kinematic centre (as determined by the model) is indicated by the white dashed line. \textbf{Right}: residual map calculated by subtracting the model PVD from the data PVD. One the left and bottom of this figure are the one-dimensional residuals created by summing over the residual map along the velocity and offset axes, respectively. The BIC of the model with radial bar-driven flows (lower row) is smaller than that of the model without radial bar-driven flows (upper row) for this object.}
    \label{fig: 8083-6101_pvd}
\end{figure*}

\par In Figures~\ref{fig: 7977-9101_model}, \ref{fig: 8616-6104_model} and \ref{fig: 8083-6101_pvd} we present examples of the models generated using the process detailed in Figure~\ref{fig: model_class}. Figure~\ref{fig: 7977-9101_model} illustrates the best-fit model of the object with MaNGA Plate IFU 7977-9101 in our un-barred sub-sample, which is a Gaussian ring with no radial bar-driven flow component (in this case, we did not find a model with radial bar-driven flows with a converged solution). In Figure~\ref{fig: 8616-6104_model}, we show the models with and without radial bar-driven flows of object 8616-6104 in our barred sub-sample. This object has a more complex structure, and is best fit with a central exponential disc and outer Gaussian ring. Adding a radial bar-driven flow component also improves the fit to the data by lowering $\rm {BIC}_{w/o\ flow} = 4544$ to $\rm {BIC}_{w/\ flow} = 4488$. We overlay the spectrum of the model with radial bar-driven flows over that of the model with no radial bar-driven flow in Figure~\ref{fig: 8616-6104_model} to illustrate the improvement. 

\par In Figure~\ref{fig: 8083-6101_pvd}, we present the position-velocity diagrams (PVDs) extracted from the models with and without radial bar-driven flows of object 8083-6101 in our edge-on sub-sample (we choose an edge-on object here as kinematic features are more easy to discern in objects that are more inclined). Again, the best-fit model of this object is determined to be two Gaussian rings with radial bar-driven flows. Both the shape of the PVD extracted from the model and the residuals when compared to the data are improved by including radial bar-driven flows. In total, 19 objects in ALMaQUEST are classed as having radial bar-driven flows, which we will refer to as the ``radial-flow subset'' for the remainder of this paper. This subset is also identified in Figure~\ref{fig: optical}.


\subsection{Statistical presence of radial bar-driven flows} 
\label{subsec: bars and gas}

\begin{figure}
    \centering
    \includegraphics[width=\linewidth]{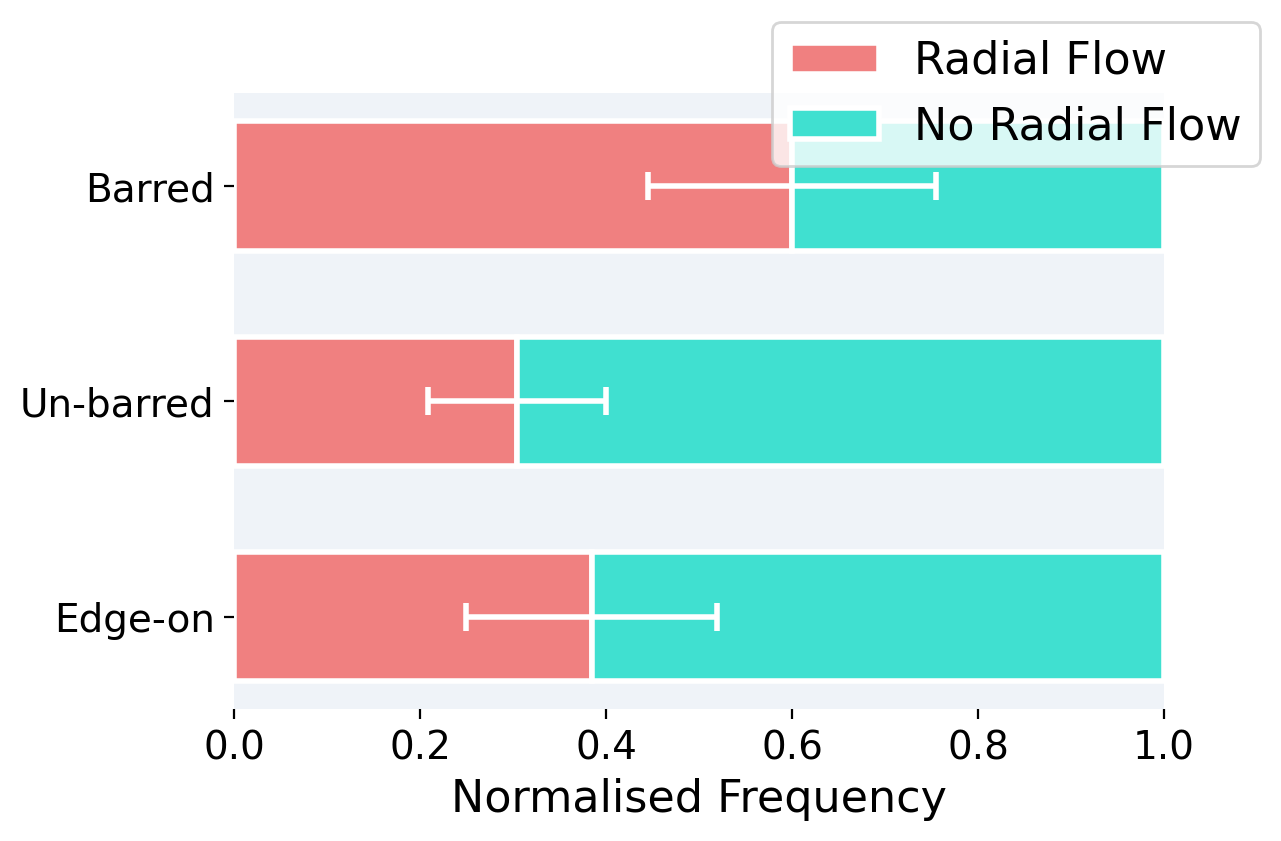}
     \caption{Bar plot representing the fraction of galaxies in each morphological sub-sample of ALMaQUEST that are determined to have radial bar-driven flows in their best-fit model. The frequency is normalised by the total number of galaxies in each sub-sample for ease of comparison. $0.60 \pm 0.15$ of the barred sub-sample objects are determined to have radial bar-driven flows, compared to $0.30 \pm 0.10$ and $0.38 \pm 0.13$ of the un-barred and edge-on sub-sample galaxies, respectively. The error bars represent the uncertainty of the radial bar-driven flow fraction of each morphological sub-sample, assuming a binomial distribution.}
     \label{fig: bar_rad_frac}
\end{figure}

Our first question, as laid out in Section~\ref{sec: intro}, requires us to determine whether we see a statistical relationship between the presence of an optical bar and the detection of non-circular kinematics like radial bar-driven flows. In Figure~\ref{fig: bar_rad_frac}, we show the proportion of galaxies in the radial bar-driven flow subset when divided into the morphological sub-samples. We find that $0.60 \pm 0.15$ of the objects in the barred sub-sample are also in the radial bar-driven flow subset, compared to $0.30 \pm 0.10$ and $0.38 \pm 0.13$ of the un-barred and edge-on sub-samples, respectively. It can be assumed that the edge-on sub-sample is composed of both barred and un-barred galaxies, so a radial bar-driven flow proportion that falls between that of the barred and un-barred sub-samples is consistent. With our bar fraction of $0.30 \pm 0.08$, and given the radial bar-driven flow detection proportions of the barred and un-barred sub-samples, we would expect $\approx 0.39$ of the edge-on sub-sample to be in the radial bar-driven flow subset, very close to our measured edge-on radial bar-driven flows fraction of $0.38 \pm 0.13$. For completion, we demonstrate the effect of small number statistics in Figure~\ref{fig: bar_rad_frac} by also showing the 1-sigma uncertainties of each morphological sub-sample, assuming a binomial distribution (i.e. uncertainty on each bar is $\sqrt{\left(p_{\rm r,i}\ p_{\rm nr,i}\right)/N_{\rm morph,i}}$, where $N_{\rm morph,i}$ are the numbers of objects in each morphological sub-sample $\rm i$, $p_{\rm r,i}$ are the radial bar-driven flow fractions and $p_{\rm nr,i}$ are the no radial bar-driven flow fractions). Our predicted fraction of edge-on galaxies with radial bar-driven flows (i.e. calculated from the bar fraction) is well within the uncertainty of this morphological sub-sample. We also note that radial motions are more easily identified in galaxies with higher inclinations, as the component of the motions projected along the line of sight will be higher. This could potentially increase the radial bar-driven flow proportion of the edge-on sub-sample, however, given how close our predicted radial bar-driven flow fraction of the edge-on sample (inferred from the barred and un-barred sub-samples) is to our measured value, this effect does not appear to be significant.

\begin{figure}
    \centering
    \includegraphics[width=\linewidth]{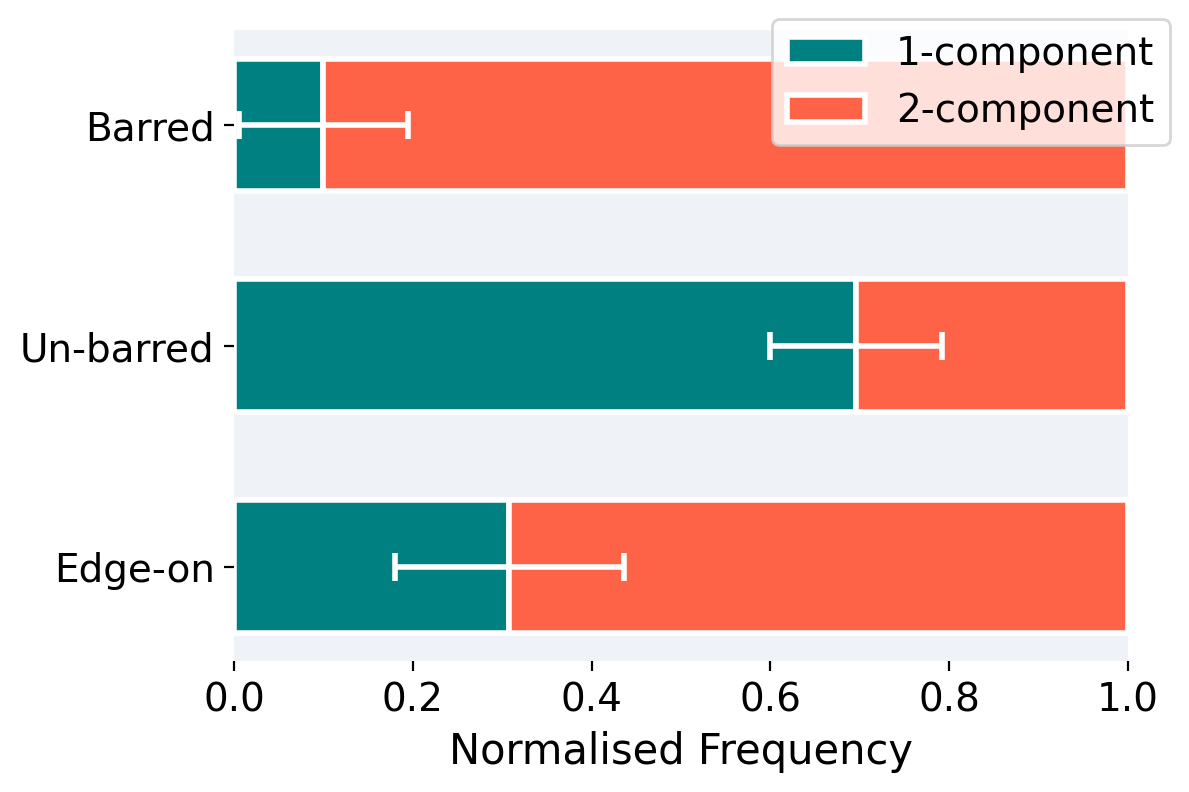}
     \caption{Bar plot representing the fraction of galaxies in each morphological sub-sample of ALMaQUEST that are determined to have a single-component (1-component) or double-component (2-component) surface brightness profile in their best-fit model (see Section~\ref{subsec: models} for more details). The frequency is normalised to the total number of galaxies in each sub-sample for ease of comparison. $0.90 \pm 0.09$ of the barred sub-sample objects are determined to have 2-component surface brightness models, compared to $0.30 \pm 0.10$ and $0.69 \pm 0.13$ of the un-barred and edge-on sub-sample galaxies, respectively. The error bars represent the uncertainty of the radial bar-driven flow fraction of each morphological sub-sample, assuming a binomial distribution.}
     \label{fig: morph_frac}
\end{figure}

\begin{figure}
    \centering
    \includegraphics[width=\linewidth]{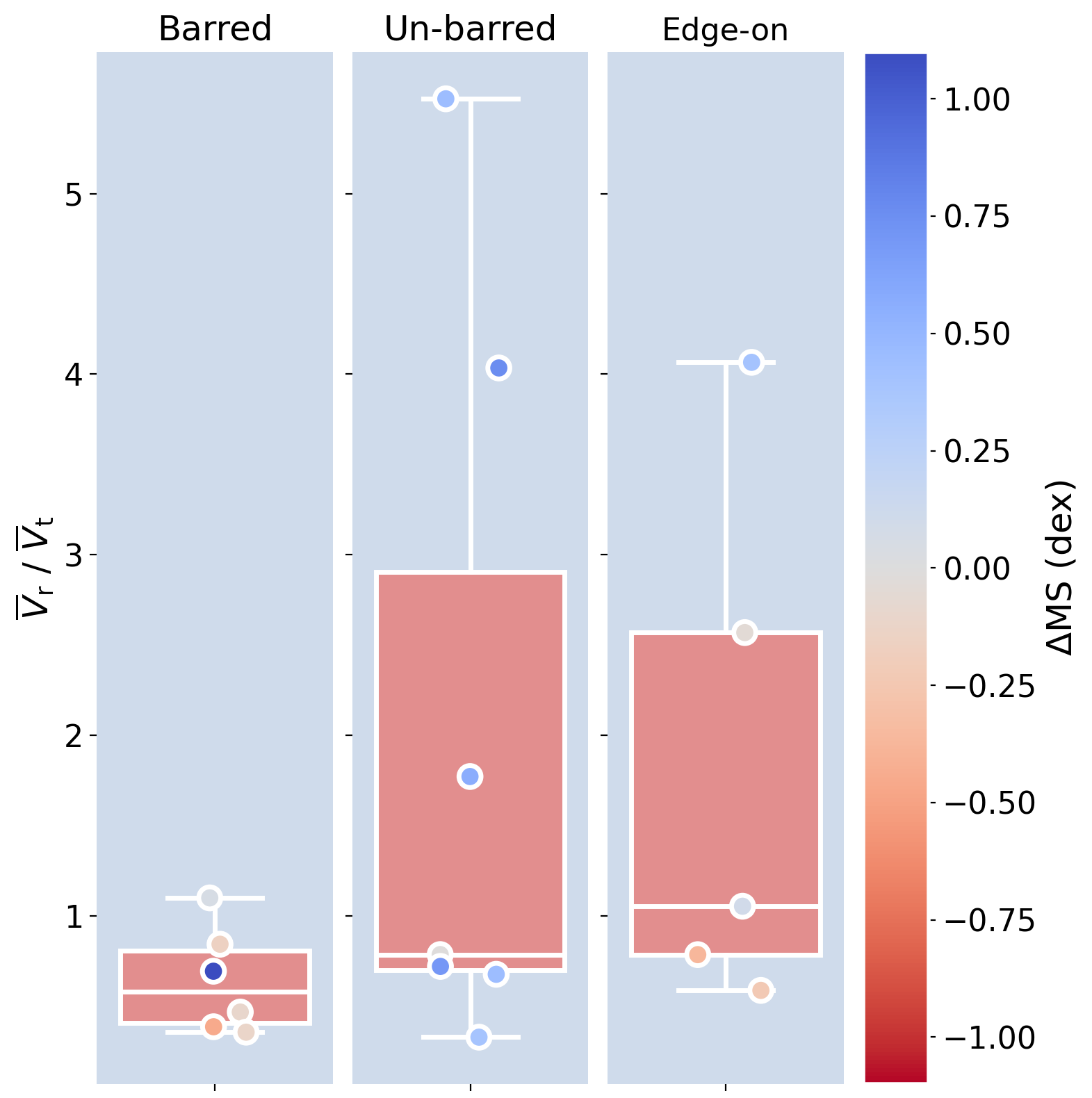}
     \caption{Distribution of the $\overline{V}_{\rm r} / \overline{V}_{\rm t}$ ratio (see text for definition) of each of the morphological sub-samples. The median position, inner quartiles and span of each distribution are shown by a box plot in each panel and the individual galaxies are shown as markers. The markers are colour-coded by their SFR offsets from the star-forming main sequence ($\rm \Delta MS$; defined in Equation~\ref{eq: delms}).}
     \label{fig: vel_ratio}
\end{figure}

\begin{figure*}
    \centering
    \includegraphics[width=0.47\linewidth]{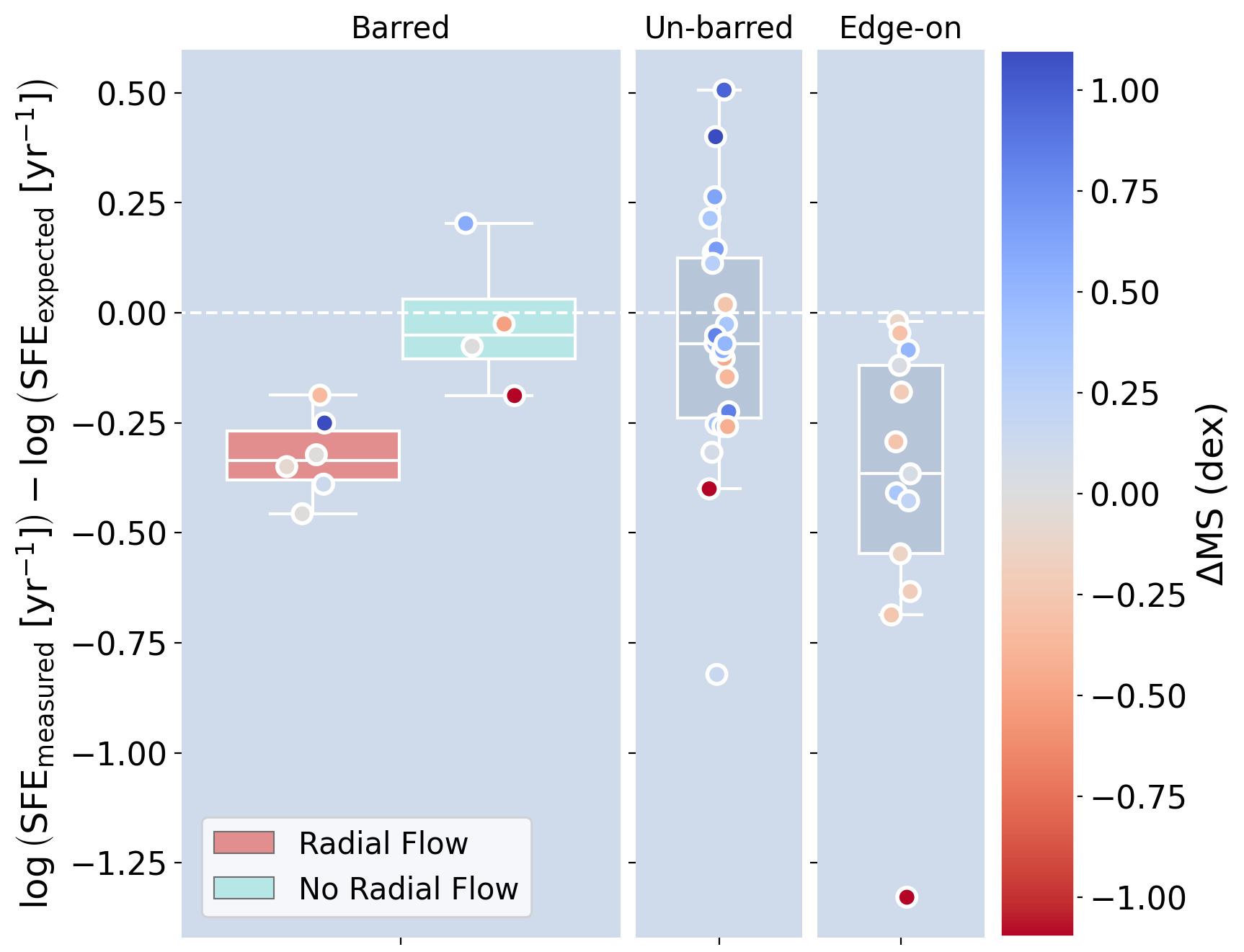}\hspace{10pt}
    \centering
    \includegraphics[width=0.4\linewidth]{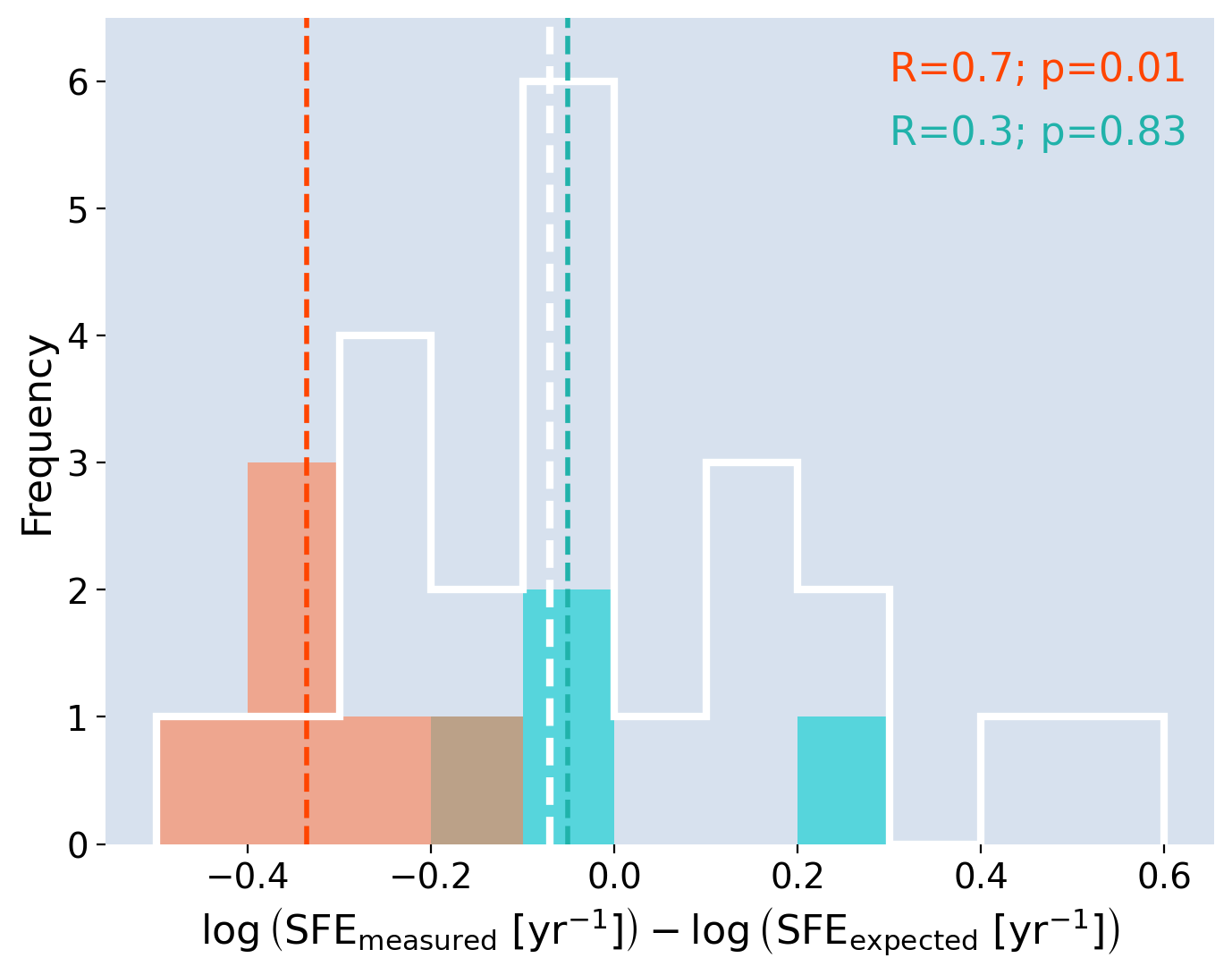}
    \caption{Distribution of the $\rm {SFE}_{measured} - {SFE}_{expected}$ quantities (see text for definition) of each of the morphological sub-samples. \textbf{Left}: the sample is divided into the three morphological sub-samples as in Figure~\ref{fig: vel_ratio}, and is split in the barred panel into the radial bar-driven flow and no radial bar-driven flow subsets. The median position, inner quartiles and span of each distribution are shown by a box plot in each panel and the individual galaxies are shown as markers. The markers are colour-coded by their SFR offsets from the star-forming main sequence ($\rm \Delta MS$). \textbf{Right}: histograms illustrating the $\rm {SFE}_{measured} - {SFE}_{expected}$ quantity for the un-barred sub-sample (white histogram) and the ``barred + radial flow'' and ``barred + no radial flow'' subsets (red and turquoise histograms, respectively). The median position of each histogram is indicated by the white, red and turquoise dashed lines for the un-barred, ''barred + radial flow'' and ``barred + no radial flow'' distributions, respectively. The 2-sided KS statistic and p-value are listed in the plot to determine whether the ''barred + radial flow'' and ``barred + no radial flow'' distributions are potentially drawn from the same distribution as that of the un-barred sub-sample. The p-value of the ''barred + radial flow'' subset suggests we can reject the null hypothesis, but we note the low numbers in our statistical analysis.}
    \label{fig: del_sfe}
\end{figure*}

\begin{figure}
    \centering
    \includegraphics[width=\linewidth]{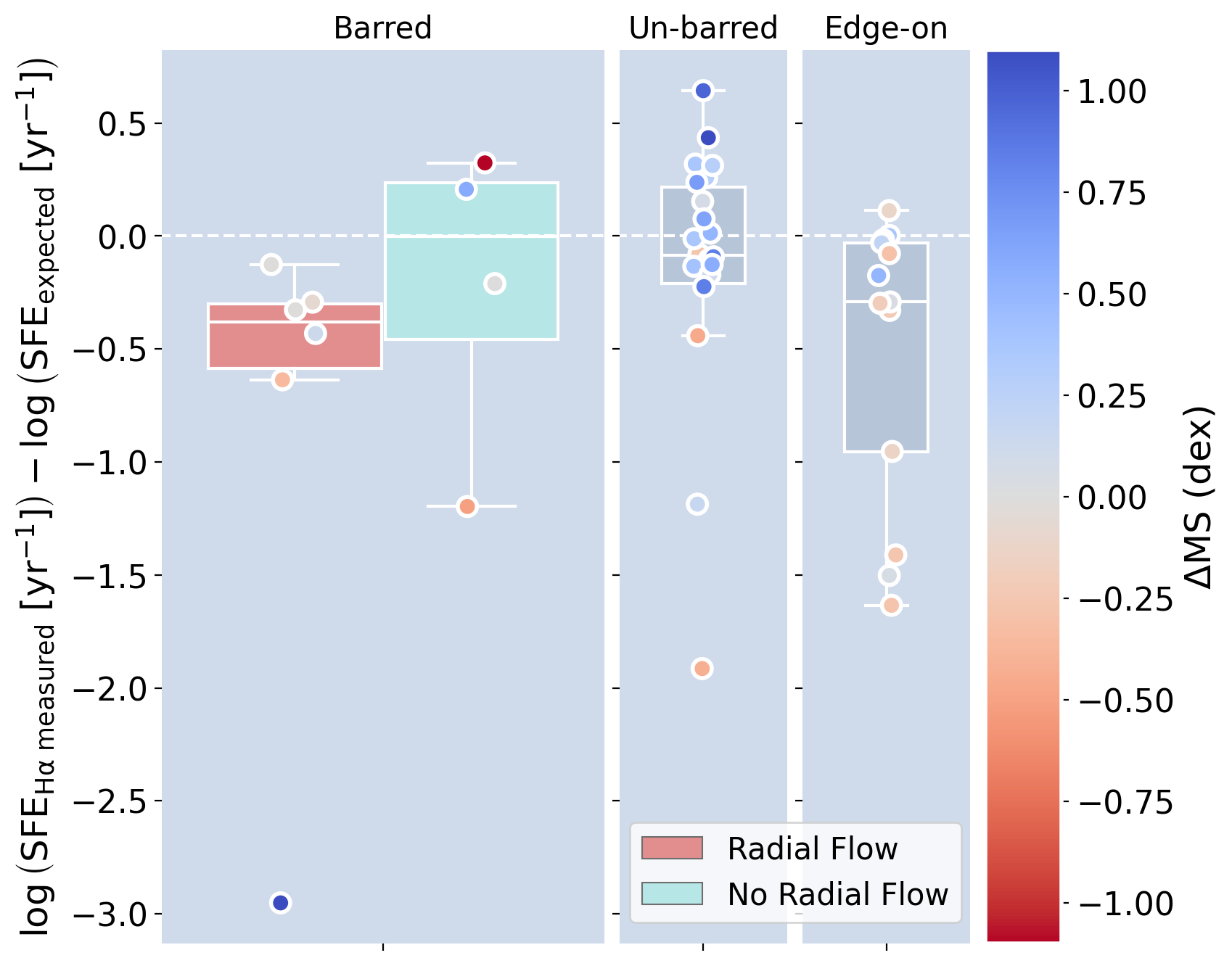}
     \caption{Distribution of the $\rm {SFE}_{H\upalpha, measured} - {SFE}_{expected}$ quantities (see text for definition) in the disc regions of each of the morphological sub-samples, using H$\upalpha$ emission from MaNGA. For each object, we sum SFR and $\rm M_{H2}$ spaxels within a circular annulus between 1~kpc and $\rm r_{50,{\it r}}$ in order to estimate the total disc SFEs. The sample is divided into the three morphological sub-samples as in Figure~\ref{fig: del_sfe} and is split within the barred panel into radial bar-driven flow and no radial bar-driven flow subsets. The median position, inner quartiles and span of each distribution are shown by a box plot in each panel and the individual galaxies are shown as markers. The markers are colour-coded by their SFR offsets from the star-forming main sequence ($\rm \Delta MS$).}
     \label{fig: sfe_disc}
\end{figure}

\par In Figure~\ref{fig: morph_frac}, we look at the proportion of galaxies that require a 1- or 2-component surface brightness model in their best-fit \textsc{KinMS} model (see Section~\ref{subsec: models}) for each morphological sub-sample. The top panel in the figure illustrates that the barred sub-sample is dominated by 2-component models ($0.90 \pm 0.09$), more so than the un-barred and edge-on sub-samples ($0.30 \pm 0.10$ and $0.69 \pm 0.13$, respectively). 2-component surface brightness models are generally associated with multiple gas rings, such as those caused by the forcing frequency of an axi-asymmetric feature like a bar. It follows, therefore, that our barred sub-sample would contain a high fraction of multiple-component models. We also anticipate that the fraction of barred sub-sample objects with multi-component models would be greater than the fraction with radial bar-driven flows, as flows are transient while resonant structures exist for far longer timescales.

\par Using the objects in our radial bar-driven flow subset, in Figure~\ref{fig: vel_ratio} we look at the distribution of the ratios of the radial velocity components ($\overline{V}_{\rm r}$) and the transverse velocity components ($\overline{V}_{\rm t}$) of the radial bar-driven flow models (see Section~\ref{subsec: models}). If $\overline{V}_{\rm r}\ /\ \overline{V}_{\rm t} < 1$, the gas is more dominated by the transverse velocity component and, therefore, is spiraling inwards as opposed to if $\overline{V}_{\rm r}\ /\ \overline{V}_{\rm t} > 1$, which would indicate that the gas is moving more directly in the radial direction. In Figure \ref{fig: vel_ratio}, we observe that the $\overline{V}_{\rm r}\ /\ \overline{V}_{\rm t}$ ratios of our barred sub-sample are all $\lesssim$ 1, whereas we see a greater spread of ratios in both the un-barred and edge-on sub-samples. This could imply that the un-barred and edge-on sub-samples contain objects undergoing a range of dynamical processes driving the flows of molecular gas, as opposed to the barred sub-sample where we see more consistency. We can infer from this that there may be a similar kinematic process driving the flows of gas in the barred sub-sample (e.g. the forcing frequency of a bar).


\subsection{Effect of radial bar-driven flows on the global SFEs of barred galaxies}
\label{subsec: SFE}

\begin{figure*}
    \centering
    \includegraphics[width=0.47\linewidth]{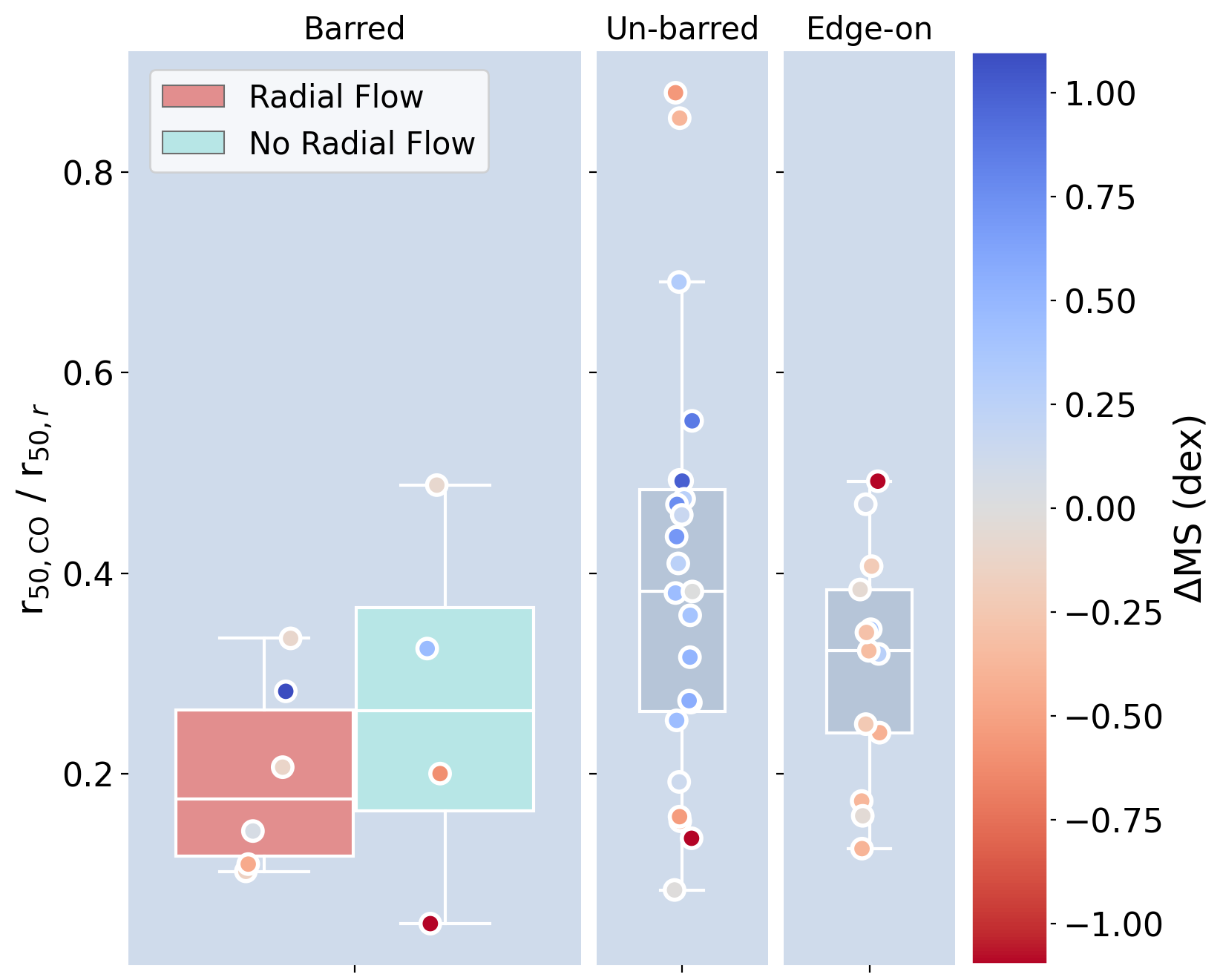}\hspace{10pt}
    \centering
    \includegraphics[width=0.4\linewidth]{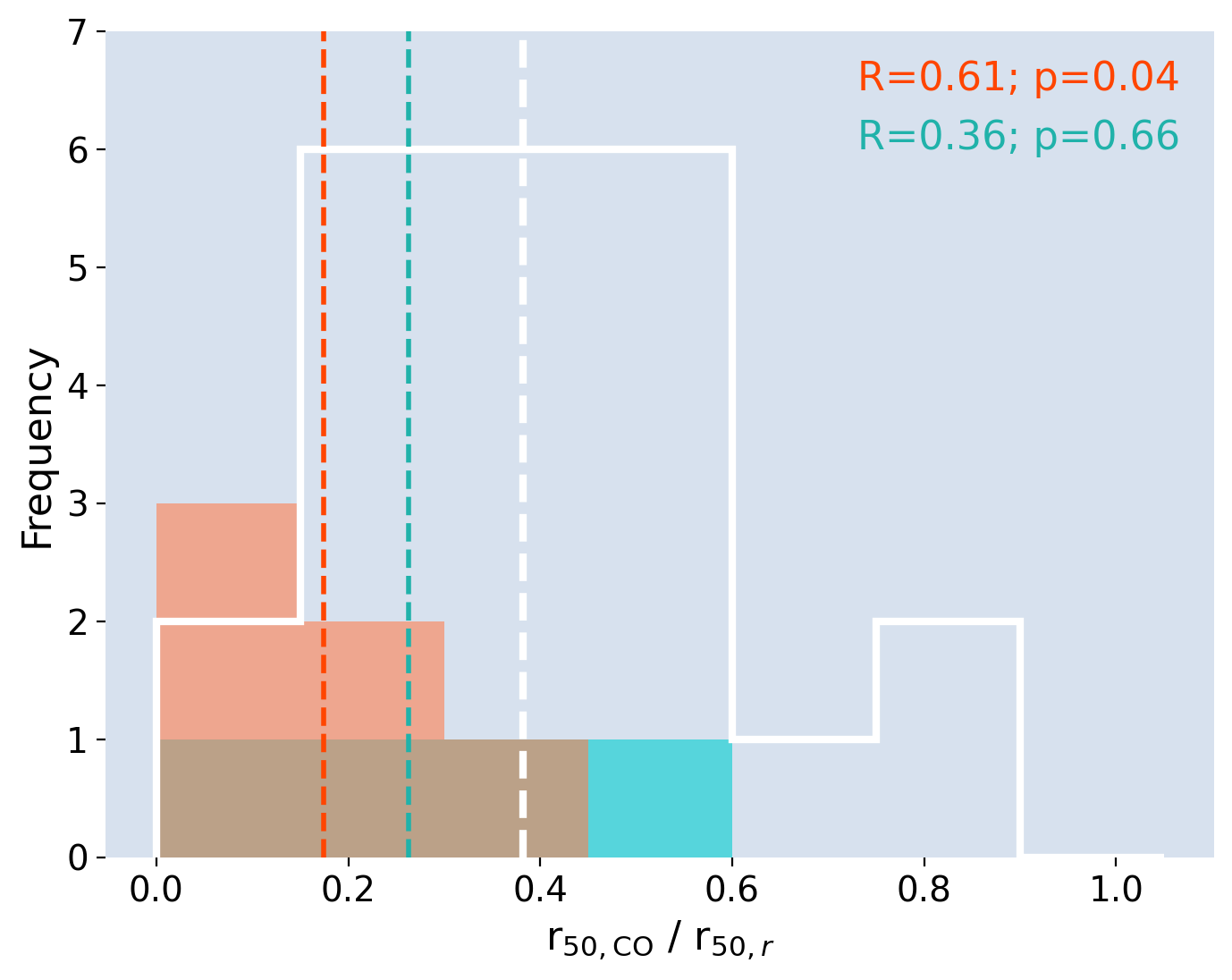}
    \caption{Distribution of the concentration parameters ($\rm r_{50,CO} / r_{50,{\it r}}$\; see text for more details) of each of the morphological sub-samples. \textbf{Left}: the sample is divided into the three morphological sub-samples as in Figures~\ref{fig: vel_ratio}-\ref{fig: sfe_disc}, and is split within the barred panel into the radial bar-driven-flow and no radial bar-driven flow subsets. The median position, inner quartiles and span of each distribution are shown by a box plot in each panel and the individual galaxies are shown as markers. The markers are colour-coded by their SFR offsets from the star-forming main sequence ($\rm \Delta MS$). \textbf{Right}: histograms illustrating the concentration parameters of the un-barred sub-sample (white histogram) and the ``barred + radial flow'' and ``barred + no radial flow'' subsets (red and turquoise histograms, respectively). The median position of each histogram is indicated by the white, red and turquoise dashed lines for the un-barred, ''barred + radial flow'' and ``barred + no radial flow'' distributions, respectively. The 2-sided KS statistic and p-value are listed in the plot to determine whether the ''barred + radial flow'' and ``barred + no radial flow'' distributions are potentially drawn from the same distribution as that of the un-barred sub-sample. Both p-values suggest that we cannot reject the null hypothesis and that the ``barred + radial flow'' subset is likely drawn from the same distribution as that of the un-barred sub-sample.}
     \label{fig: cons}
\end{figure*}

\begin{figure}
    \centering
    \includegraphics[width=\linewidth]{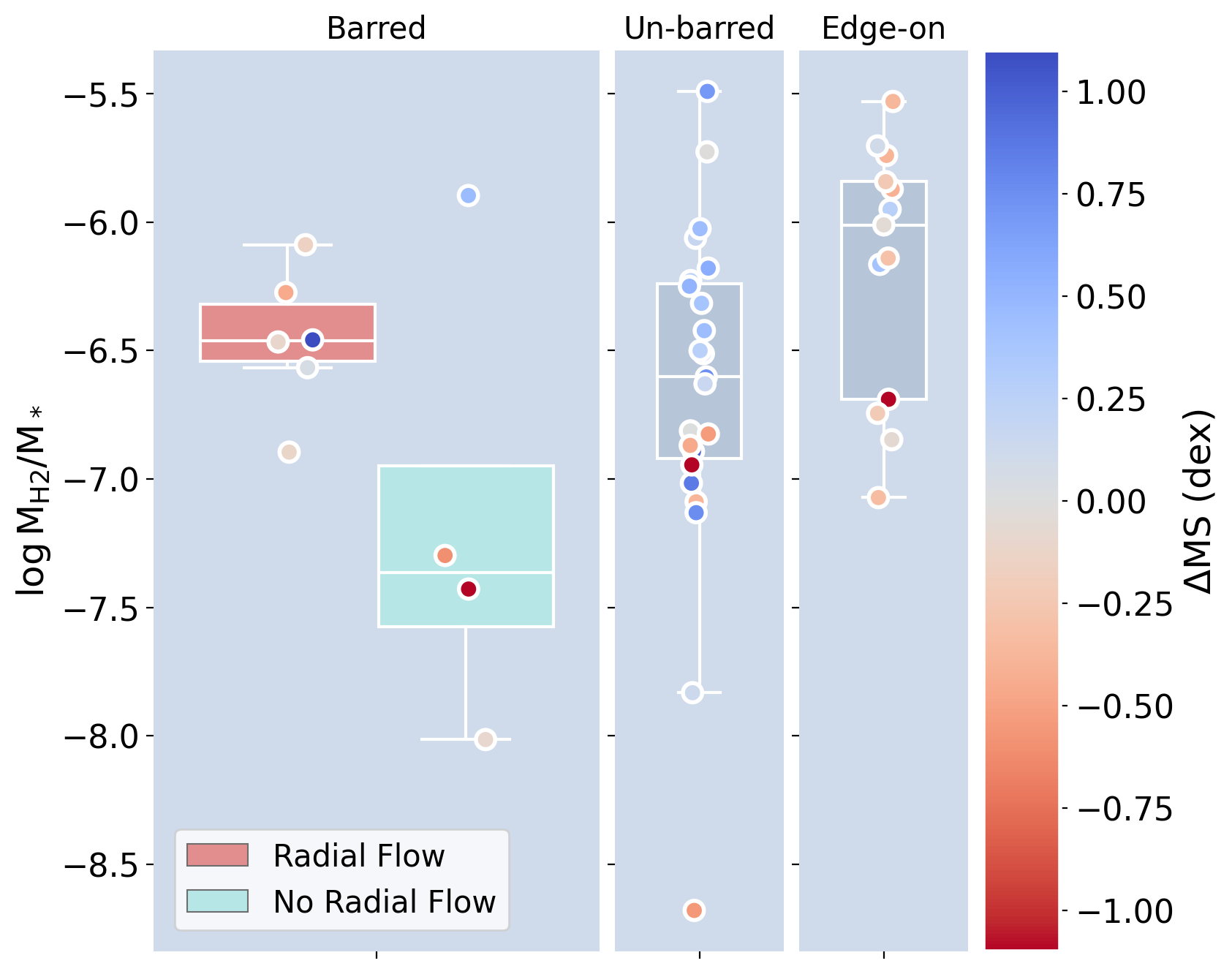}
   
     \caption{Distribution of the molecular gas fractions ($\rm M_{H2} / M_*$) in the central 1~kpc radius of each of the morphological sub-samples, using stellar-mass maps from MaNGA (see Section~\ref{subsec: manga}). The sample is divided into the three morphological sub-samples as in Figures~\ref{fig: del_sfe}-\ref{fig: cons} and is split within the barred panel into radial bar-driven flow and no radial bar-driven flow subsets. The median position, inner quartiles and span of each distribution are shown by a box plot in each panel and the individual galaxies are shown as markers. The markers are colour-coded by their SFR offsets from the star-forming main sequence ($\rm \Delta MS$).}
     \label{fig: gas_dists}
\end{figure}

\begin{figure*}
     \centering
    \includegraphics[width=0.47\linewidth]{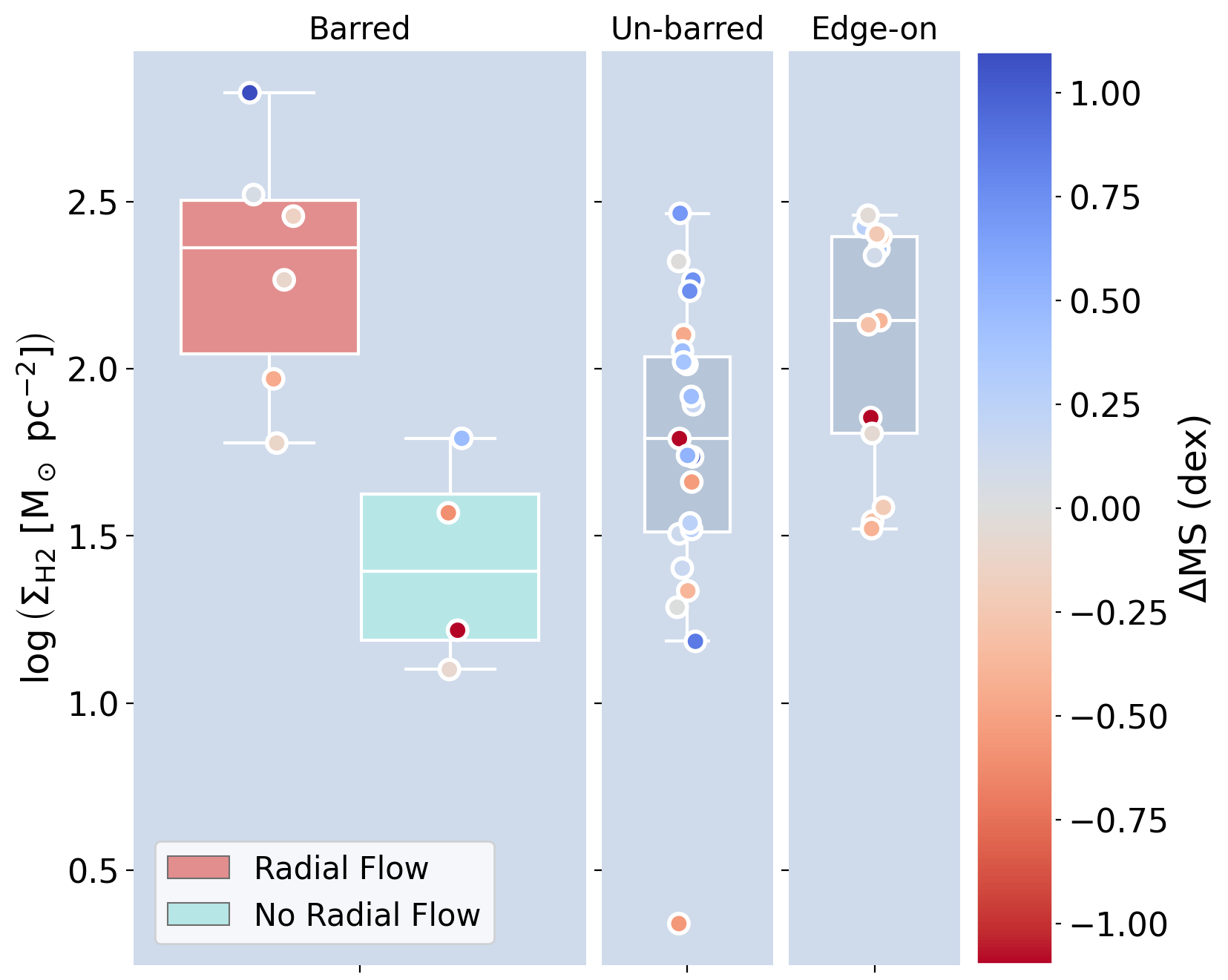}\hspace{10pt}
    \centering
    \includegraphics[width=0.47\linewidth]{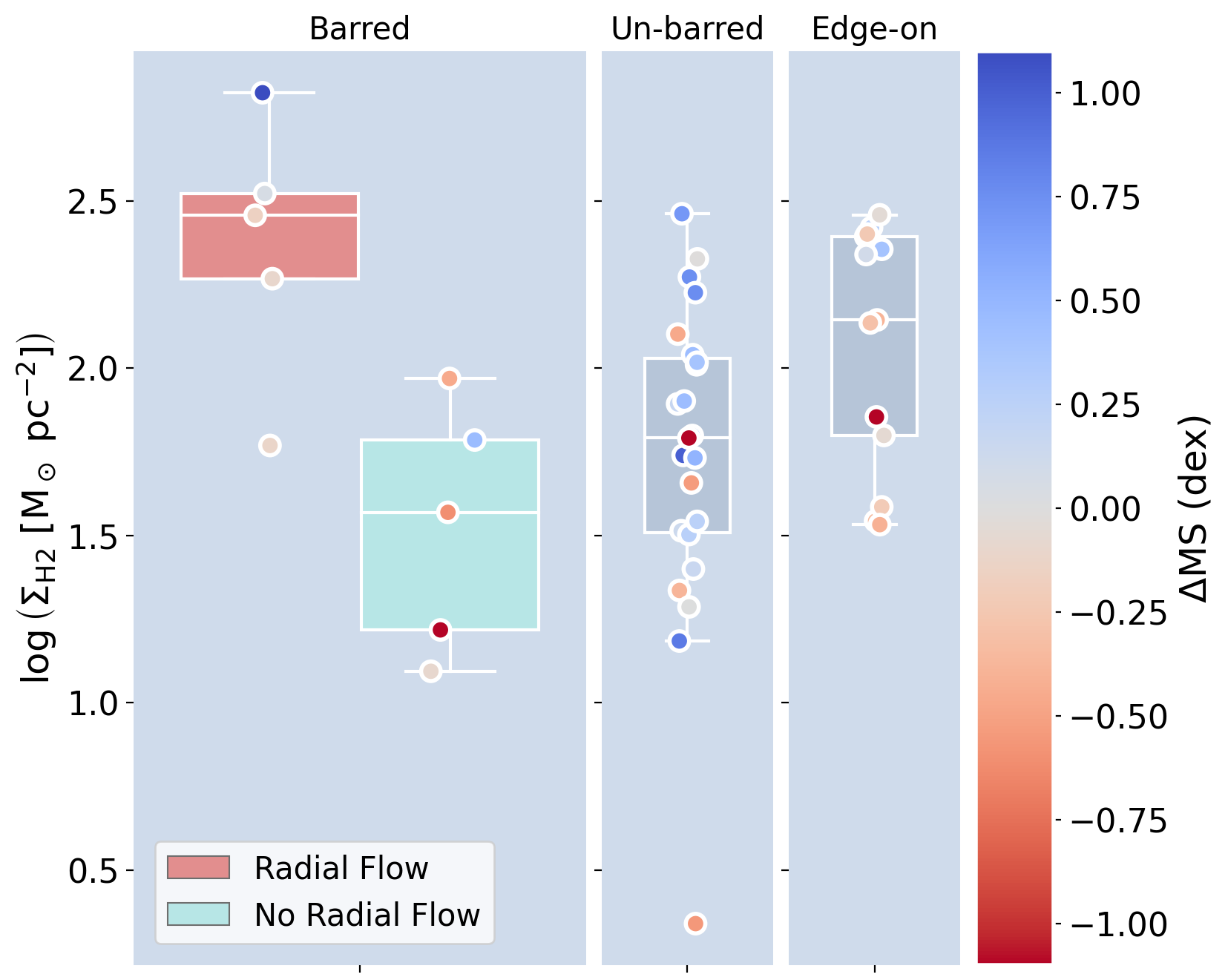}
    \caption{Distribution of the molecular gas mass surface densities ($\rm \Sigma _{H2}$) in the central 1~kpc radius of each of the morphological sub-samples. \textbf{Left}: the sample is divided into the three morphological sub-samples as in Figures~\ref{fig: vel_ratio}-\ref{fig: gas_dists}, and is split within the barred panel into the radial bar-driven-flow and no radial bar-driven flow subsets. The median position, inner quartiles and span of each distribution are shown by a box plot in each panel and the individual galaxies are shown as markers. The markers are colour-coded by their SFR offsets from the star-forming main sequence ($\rm \Delta MS$). \textbf{Right}: as the left hand side panel, but using radial flow subsets defined once the ALMaQUEST cubes have been degraded to have S/N < 30 (see text for detail).}
    \label{fig: gas_density}
\end{figure*}

Our second question posed in Section~\ref{sec: intro} involves investigating how non-circular molecular gas kinematics affects the global SFEs of barred galaxies. We define the SFE as the ratio of the total SFR ($\rm {SFR}_{tot}$) and the total H2 gas mass ($\rm {M}_{H2,tot}$). The $\rm {SFRs}_{tot}$ we use are detailed in Sections~\ref{subsubsec: gswlc} and \ref{subsubsec: wise} from the GSWLC-X2 and WISE catalogues, and we calculate $\rm {M}_{H2,tot}$ using our masked ALMaQUEST $^{12}$CO(1-0) datacubes (see Section~\ref{subsec: ALMaQUEST}). 

\par In Figure~\ref{fig: del_sfe} we plot the distribution of $\rm \log({SFE}_{measured}[{yr}^{-1}])$ - $\rm \log ({SFE}_{expected}[{yr}^{-1}])$, which estimates the enhancement/suppression of SFE relative to what would be expected given an object's location on the star-forming main sequence. $\rm \log ({SFE}_{measured})$ are the SFEs we calculate with $\rm {SFR}_{tot}$ and $\rm {M}_{H2,tot}$. $\rm \log ({SFE}_{expected})$ is a function of the object's offset from the star-forming main sequence ($\rm \Delta MS~[dex]$), which we define as: 

\begin{equation}
    \begin{aligned}
        \rm \Delta MS~[dex] = \log \left(SFR_{tot}~\left[M_\odot ~{yr}^{-1}\right]\right) - \\ \rm \log \left( SFR_{MS}(M_* / M_\odot)~\left[M_\odot ~{yr}^{-1}\right]\right)\ ,
    \end{aligned}
    \label{eq: delms}
\end{equation}

\hspace{2pt} where $\rm SFR_{MS}~(M_*)$ is the star-forming main sequence as a function of stellar mass ($\rm M_*$) using \citet{saintonge22} function, which is revised from the \citet{saintonge16} using $\rm M_*$ and SFRs from GSWLC-2: 

\begin{equation}
    \begin{aligned}
        \rm \log \left( {SFR}_{MS}~(M_*)~\left[M_\odot ~{yr}^{-1}\right]\right) = 0.412 - \\ \rm \log \left( 1 + \left[ \frac{10^{M_* / M_\odot}}{10^{10.59}} \right] ^{-0.718} \right) \ .
    \end{aligned}
    \label{eq: sfrms}
\end{equation}

We then derive $\rm \log ({SFE}_{expected}(\Delta MS))$ by fitting a line to the $\rm \Delta \log (SFE)$ illustrated in Figure~9 of \citet{saintonge22}, with the form:

\begin{equation}
    \begin{aligned}
        \rm \log \left( {SFE}_{expected}~(\Delta MS)~\left[{yr}^{-1}\right]\right) = 0.29 \times \Delta MS + \\ \rm 0.04 + (-9.12)\ ,
    \end{aligned}
    \label{eq: sfrexp}
\end{equation}

\hspace{2pt} where -9.12 is the median $\rm \log (SFE)$ of ALMaQUEST galaxies in the un-barred sub-sample that are within $\pm 0.3$ dex of the star-forming main sequence (this aids interpretation of the relative offset between samples with respect to un-barred main sequence galaxies). Any enhancement/suppression of SFE measured for our barred sub-sample galaxies, therefore, may purely be a consequence of their locations relative to the star-forming main sequence. Equally, the presence of a bar could be driving the position of a galaxy on the main sequence. For both scenarios, using $\rm \log({SFE}_{measured}) - \log ({SFE}_{expected})$ should reduce this effect.

In the left hand side panel of Figure~\ref{fig: del_sfe}, we see an indication that the ``barred + radial flow'' subset galaxies have generally suppressed SFEs, relative to both the ``barred + no radial flow'' subset and un-barred sub-sample galaxies. We note that these subsets are small, particularly the ``barred + no radial flow'' subset, so our analysis is once again limited by the small number of galaxies available in this work. However, we see a tenuous suggestion that the presence of radial flows in barred galaxies suppresses their SFEs relative to the ``barred + no radial flow'' subset and un-barred sub-sample. In the right hand side panel of Figure~\ref{fig: del_sfe}, we plot the un-barred sub-sample alongside the ``barred + radial flow'' and ``barred + no radial flow'' subsets. We calculate the 2-sided Kolmogorov-Smirnov (KS) statistic for both radial flow barred subsets with the un-barred sub-sample and find that while the ``barred + no radial flow'' subset is likely drawn from the same distribution as the un-barred sub-sample, the p-value of the ``barred + radial flow'' subset is < 0.05. This implies that we can reject the null-hypothesis at 2$\sigma$ and the ``barred + radial flow'' subset galaxies are likely drawn from a different distribution than that of the un-barred sub-sample.   

\par Using SFR maps derived from H$\upalpha$ emission from MaNGA, in Figure~\ref{fig: sfe_disc} we plot the distributions of $\rm {SFE}_{H\upalpha,measured} - {SFE}_{expected}$ for the disc regions of the ALMaQUEST galaxies. We exclude the central region due to some objects (particularly in the barred sub-sample) having this area masked due to AGN-contamination or low-S/N. With the SFR maps calculated using the method detailed in Section~\ref{subsec: manga}, we estimate the total disc SFR by summing spaxels in a circular annulus between 1~kpc and the $r$-band effective radius from the NSA catalogue for each object ($\rm r_{50,{\it r}}$). $\rm M_{H2}$ spaxels then are summed in the same region and used to calculate $\rm {SFE}_{H\upalpha,measured}$. The $\rm {SFE}_{expected}$ values used are the same as those derived in Equations~\ref{eq: delms}-\ref{eq: sfrexp} using $\rm {SFR}_{tot}$ from GSWLC-X2 and WISE. We find a similar relative distribution between our morphological sub-samples as we observe in Figure~\ref{fig: del_sfe}. Again, we are limited in particular when we interpret differences between the radial flow subsets in this figure due to the size of the samples, but we do see an indication that the median value of the ``barred + radial flow'' distribution is suppressed relative to that of the ``barred + no radial flow'' subset, which is approximately coincident with that of the un-barred sub-sample (although there is significant scatter). Using the same method as in Figure~\ref{fig: del_sfe}, we use the 2-sided KS test to discern whether the SFE values from the radial bar-driven flows subsets in the barred sub-sample are drawn from the same distribution as the un-barred sub-sample. We find again that the p-value < 0.05 for the ``barred + radial flow'' subset but >> 0.05 for the ``barred + no radial flow'' subset. This could imply that SFE is suppressed in the discs of galaxies in the ``barred + radial flow'' subset relative to those in the unbarred sub-sample, which may contribute to the variations in global SFE distributions that we observe in Figure~\ref{fig: del_sfe}. In Appendix~\ref{appendix: sfe}, we also attempt to observe differences in $\rm {SFE}_{H\upalpha,measured} - {SFE}_{expected}$ between our morphological sub-samples and radial flow subsets in the central 1~kpc using median SFE spaxel values available in these regions.
 
\par We conduct a similar analysis to that carried out in Figure~\ref{fig: del_sfe} in Figure~\ref{fig: cons}, where we look at the distributions of concentration parameters of the objects in our three morphological sub-samples. The concentration parameter is defined as the half-light radius of the CO(1-0) intensity of each object, divided by $r_{50,{\it r}}$ (i.e. $\rm r_{50,CO}/ r_{50,{\it r}}$). In the left hand side panel of Figure~\ref{fig: cons}, we show the distribution of concentration parameters of each of the morphological sub-samples and the barred radial flow subsets. The objects in the ``barred + radial flow'' subset all have $\rm r_{50,CO} / r_{\rm 50,{\it r}} \lesssim 0.4$, whereas in the other morphological sub-samples there are broader distributions of concentration parameters. In the right hand side panel of Figure~\ref{fig: cons} we compare the un-barred concentration parameter distributions with those of the barred sub-sample's radial flow subsets (in the same manner as the corresponding plot in Figure~\ref{fig: del_sfe}). Again in this analysis, we find that the p-value of the distribution in the ``barred + radial flow'' subset is < 0.05, suggesting we can reject the null hypothesis, while that for the ``barred + no radial flow'' subset is >> 0.05. This continues to suggest a similarity in the distributions of the ``barred + no radial flow'' subset and un-barred sub-sample, with the ``barred + radial flow'' subset appearing to be statistically distinct.  

\par As addressed in Section~\ref{subsec: manga}, spatially-resolved analyses of SFRs in ALMaQUEST are limited by AGN contamination and/or low S/N in the central regions of some of the galaxies in the sample. However, in Figures~\ref{fig: gas_dists} and \ref{fig: gas_density}, we look at the molecular gas fraction and mass surface density within the central 1~kpc radius of each galaxy. In order to calculate the gas fractions ($\rm \log M_{H2} / M_*$), we use the stellar-mass surface density maps from MaNGA (see Section~\ref{subsec: manga}). Using the same method detailed previously, we calculate the 2-sided KS statistic of the central gas fractions of the ``barred + radial flow'' and ``barred + no radial flow'' subsets shown in Figure~\ref{fig: gas_dists} with those of the un-barred sub-sample. We find that the p-value in each case is $>0.05$, suggesting that both barred radial flow subsets have central gas fractions drawn from the same distribution as that of the un-barred sub-sample. Using the same analysis, in the left hand side panel of Figure~\ref{fig: gas_density}, we show the distribution of molecular gas mass surface densities ($\rm \log \Sigma _{H2}$) of each of the morphological sub-samples. The median of the central molecular gas mass surface densities of the ``barred + radial flow'' subset is clearly enhanced relative to that of the ``barred + no radial flow'' subset and un-barred sub-sample. Using the 2-sided KS statistic, we find that the null hypothesis can be rejected for the ``barred + radial flow'' subset, but not for the ``barred + no radial flow'', when compared to the un-barred sub-sample. This may imply that the centres of galaxies hosting radial bar-driven flows are generally more gas-rich.

\par In Figure~\ref{fig: gas_density}, we consider whether the variations of S/N ratios of the ALMaQUEST molecular gas maps could explain the enhanced central molecular gas mass surface densities observed in the ``barred + radial flow'' subset. We provide more detail of this process in Appendix~\ref{appendix: snr}, where we show that our results are robust against degrading all the original ALMaQUEST CO(1-0) datacubes to S/N < 30. The right hand side panel of Figure~\ref{fig: gas_density} shows the molecular gas mass surface densities when all the ALMaQUEST datacubes have S/N < 30. The main difference in this panel is that one object changes subset from the ``barred + radial flow'' subset to the ``barred + no radial flow'' subset when we use \textsc{KinMS} to re-model the degraded datacubes. However, the median position of the central molecular gas mass surface densities from the ``barred + radial flow'' subset remains higher than that of the ``barred + no radial flow'' subset and the results from the 2-sided KS test remain consistent with those calculated in the left hand side panel.

\section{Discussion}
\label{sec: discussion}

\subsection{Features of radial bar-driven flows:}
\label{subsec: gas flow}

\subsubsection{Barred galaxies}
\label{subsec: bar gas flow}

The primary question addressed in this work is whether there is a statistical relationship between the presence of an optical bar in a galaxy and the detection of non-circular kinematics of molecular gas (with a focus on radial bar-driven flows). We find that $0.60 \pm 0.15$ of galaxies in ALMaQUEST that we classify as barred (see Section~\ref{subsec: morph class} for details of the classification system) also host radial molecular gas flows (see Section~\ref{subsec: models} for our definition of radial bar-driven flows). This is in comparison to the $0.30 \pm 0.10$ and $0.38 \pm 0.13$ \textbf{of} the un-barred and edge-on sub-samples, respectively, which we illustrate in Figure~\ref{fig: bar_rad_frac} (where the edge-on sub-sample is composed of both barred and un-barred galaxies). We can infer from this that barred galaxies are statistically more likely to possess non-circular molecular gas kinematics that resembles radial bar-driven flows \citep[i.e. gas moving in the plane of a galaxy as the result of the forcing frequency of a bar;][]{spekkens07}. This finding echoes numerical simulations of molecular gas transport in barred galaxies, such as those of \citet{krumholtz15}, where the role of large-scale galactic structures like bars are emphasised as critical drivers of gas dynamics and star formation at the centres of galaxies. However, \citet{kruijssen14} and \citet{krumholtz15} also highlight the importance of timescale when observing radial gas motions. Their models cycle through distinct phases of accumulation, starburst, outflow and quenching, which results in a range of potential observations, depending on the phase in which a galaxy is observed. While radial bar-driven flows are more likely in our barred sub-sample galaxies, it also follows that it may not be present in all of our barred objects due to the impact of dynamical timescales.

\par In addition to the greater likelihood of radial bar-driven flow detection in our barred sub-sample, we also find that $0.90 \pm 0.09$ of these objects are best fit by a 2-component surface brightness model (instead of a 1-component model; see Section~\ref{subsec: models} for details on the modelling process). Our un-barred and edge-on sub-samples are significantly less likely to be best-fit by 2-component surface brightness models, with $0.30 \pm 0.10$ and $0.69 \pm 0.13$ preferring a 2-component model over a 1-component model, respectively (see Figure~\ref{fig: morph_frac}). A multi-component surface brightness model can indicate the presence of a resonance system, where molecular gas is driven onto multiple resonant orbits by the pattern frequency of a bar or other axi-asymmetric feature \citep[e.g. see][]{combes91, comeron14, fraser-mckelvie20, chiba21}. Strong resonant orbits occur at co-rotation and at the Lindblad resonances, where an orbit's epicyclic frequency is commensurate with the pattern frequency of the central bar. Molecular gas rings have been directly observed in multiple previous studies, including \citet{lu22}, where the authors find evidence of a molecular gas inflow along the bar of the galaxy PGC34107, forming an asymmetric gas ring at the inner Lindblad resonance (ILR) using high-resolution millimeter interferometry from the Northern Extended Millimeter Array (NOEMA). We can infer, therefore, that a similar process may be underway in our barred sub-sample galaxies, whereby molecular gas is funneled along their bars into resonant orbits.

\par It is also noteworthy that the fraction of barred galaxies best-fit by 2-component surface brightness models is greater than the fraction in the ``barred + radial flow'' subset. As considered in Section~\ref{subsec: bars and gas}, this difference can be predicted from the literature, which generally points to radial bar-driven flows caused by bar instabilities being shorter lived than the lifetime of the bar itself. For example, \citet{sormani18} and \citet{sormani19} infer from their simulations of the CMZ that the intrinsic morphological asymmetry of a galactic bar leads to intrinsically transient and time variable gas inflows. Moreover, \citet{schinnerer23} study the inner 5~kpc of the local barred spiral galaxy NGC 1365 with the James Webb Space Telescope/Mid-Infrared Instrument (JWST/MIRI) imaging alongside ALMA CO(2-1) mapping from PHANGS-ALMA (PHANGS: Physics at High Angular resolution in Nearby GalaxieS). They find asymmetric gas distributions along the ``bar lanes'' \citep[defined in][as the distributions of gas and dust along the leading sides of bars that extend out towards the disc]{sormani18} and a lopsided star-formation distribution, as well as evidence of streaming motions along the ``bar lanes''. Using hydrodynamical simulations, they are able to replicate these observations, with transient streaming motions that occur within a dynamical time (which in the case of NGC 1365 is $\approx 30$~Myr, with streaming motion timescales as small as $\approx 6$~Myr). Largely, this is attributed to the clumpiness of the gas distribution, which subsequently forms a highly variable inflow along the bar. \citet{sormani23} find very similar results when studying NGC 1097, which hosts cold gas inflow rates that vary over timescales of $\approx$ 10~Myr. Collectively, these studies strongly support a scenario of episodic gas accretion through short-lived radial inflows. When comparing these inflow timescales to the lifetime of a bar, the literature suggests that a bar can be stable over a period of 1-10 Gyr \citep[e.g.][]{shen_sellwood04, gadotti06, gadotti15, cavanagh22}. It follows, therefore, that the long lifetime of the bar would maintain resonant structures within a galaxy on timescales that far exceed the lifetimes of transient radial flows. Consequently, it is more likely at any one time to observe a resonant gas structure than it is to observe a radial bar-driven flow in a barred galaxy, as we find in Section~\ref{subsec: bars and gas}.

\par Another consistent property of the molecular gas in our barred sub-sample galaxies, especially the radial bar-driven flow subset, is low concentration parameters (defined as $\rm r_{50,CO} / r_{50,{\it r}}$ in Section~\ref{subsec: SFE} and illustrated in Figure~\ref{fig: cons}). We find a narrower range of concentration parameters for this subset of galaxies, at the low end of the distribution, suggesting that the gas in these objects is concentrated relative to that in the un-barred and edge-on sub-sample galaxies. The range of concentration parameters is also narrower than that measured for the objects in the barred sub-sample without radial bar-driven flow detection. However, the small size of this subset makes the comparison somewhat tenuous. This result is consistent with the findings of \citet{yu22}, who use a sample of similar size drawn from EDGE-CALIFA (EDGE: Extragalactic Database for Galaxy Evolution; CALIFA: Calar Alto Legacy Integral Field Area) to assess the correlation between central molecular gas concentration and large-scale galaxy asymmetry. In particular, the authors focus on the difference of gas concentrations between galaxies with and without an optical bar, but do not find a statistically-significant difference between the concentration parameters they calculate for these two sub-samples. This is despite the fact that their barred distribution peaks at higher gas concentrations than their their un-barred distribution. Our results illustrated in Figure~\ref{fig: cons} support a similar scenario, where the concentration parameters of our barred sub-sample galaxies do peak at a lower value than the distribution drawn from the un-barred and edge-on sub-samples (although they are not drawn from a different distribution according to the 2-sided KS statistic). Ultimately, more data are required to assess this result, but in the context of the high fraction of resonant gas structures found in our barred sub-sample, it follows that this gas would be more centrally concentrated along inner orbits \citep[e.g.][]{sakamoto99, jogee05, kuno07, combes14}.

\begin{figure}
    \centering
    \includegraphics[width=\linewidth]{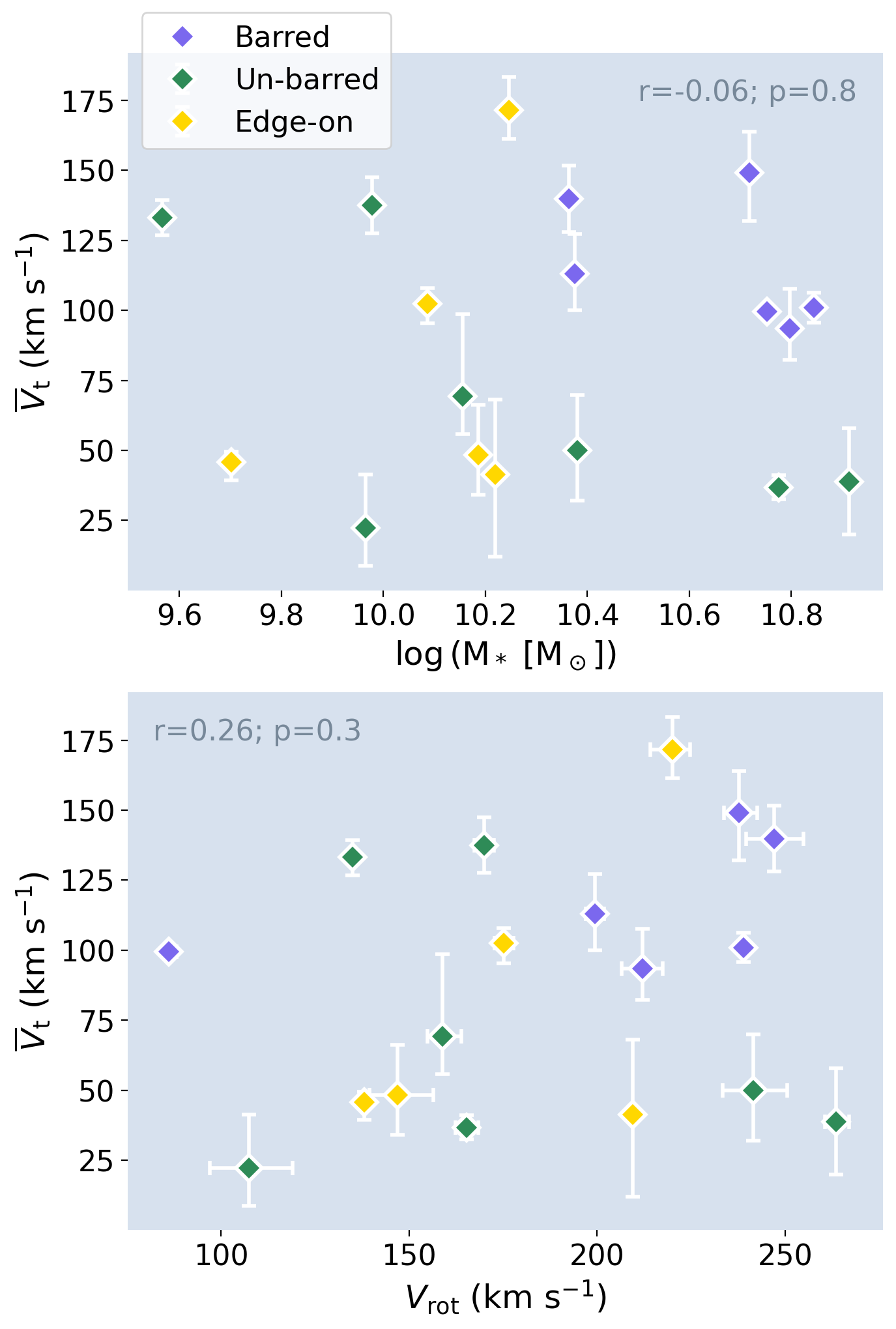}
     \caption{Dependence of the transverse velocity components ($\overline{V}_{\rm t}$) of our radial flow subsets. \textbf{Top}: plot of $\overline{V}_{\rm t}$ against $\rm M_*$. The Spearman correlation coefficient r=-0.06 and the p-value $\rm >>0.05$. These quantities are, therefore, not correlated. \textbf{Bottom}: plot of $\overline{V}_{\rm t}$ against the circular velocity ($V_{\rm rot}$). The Spearman correlation coefficient r=0.26 and the p-value $\rm>>0.05$, again suggesting there is no correlation between these quantities.}
     \label{fig: star_vr}
\end{figure}

\par We also find a notable consistency in some of the dynamical properties of the radial bar-driven flow subset of our barred sub-sample. In Figure~\ref{fig: vel_ratio}, we show that all the objects in our ``barred + radial flow'' subset have a ratio of radial to transverse velocities ($\overline{V}_{\rm r}\ /\ \overline{V}_{\rm t}$) $\lesssim 1$ (see Section~\ref{subsec: bars and gas} for more details). This is in comparison to both the un-barred and edge-on radial bar-driven flow subsets, which have a larger scatter of $ \overline{V}_{\rm r}\ /\ \overline{V}_{\rm t}$. $\overline{V}_{\rm r}\ /\ \overline{V}_{\rm t} << 1$ suggests essentially circular orbits, while $\overline{V}_{\rm r}\ /\ \overline{V}_{\rm t} >> 1$ implies significant non-circular motions. \citet{spekkens07} also comment on the relevance of the $\overline{V}_{\rm r}\ /\ \overline{V}_{\rm t}$ ratio, finding that their bisymmetric model produces $\overline{V}_{\rm r}\ /\ \overline{V}_{\rm t} = 1$ for solid-body rotation and $\overline{V}_{\rm r}\ /\ \overline{V}_{\rm t} \approx 0.67$ for flat rotation. All of the objects in our ``barred + radial flow'' subset have $R_{\rm b} > R_{\rm turn}$, meaning that their bars extend beyond the rising portion of their rotation curves (i.e. where there is roughly solid-body rotation). As our $\overline{V}_{\rm r}$ and $\overline{V}_{\rm t}$ parameters are spatially constant in the region up to $R_{\rm b}$, we are effectively averaging these velocities over the rising and flat portions of the rotation curves. Values of $\overline{V}_{\rm r}\ /\ \overline{V}_{\rm t} \lesssim 1$ for our ``barred + radial flow'' subset appear, therefore, to be in the range anticipated by \citet{spekkens07} for distortions created by the presence of a bar. It is, however, possible that our radial bar-driven flow model is capturing shocks induced along molecular ridges (i.e. long structures along the leading edges of a bar) in our barred sub-sample, which can rapidly decelerate molecular gas and potentially alter our $\overline{V}_{\rm r}$ and $\overline{V}_{\rm t}$ parameters \citep[e.g.][]{kuno00}. In the case of our un-barred and edge-on sub-samples, for which $\overline{V}_{\rm r}\ /\ \overline{V}_{\rm t} >> 1$, the measurements are outside the values anticipated by \citet{spekkens07}, reinforcing our argument that the flows in these objects are, at least in part, kinematically distinct from those in the barred sub-sample. 

\par The transverse velocity component of a bar flow should not in principle be related to the circular velocity resulting from the stellar potential. We investigate this in Figure~\ref{fig: star_vr}, where we show that there is no statistically-significant correlation (p-values $>>0.05$) between the transverse velocity components and either the total stellar masses or circular velocities of our sample galaxies. Our measured transverse velocity components, therefore, do not seem connected to the stellar potentials and can be viewed as distinct features of the gas flows of our objects. This result once again implies that there is a kinematic similarity in how cold gas is moving in our barred sub-sample galaxies, compounding the evidence that they are undergoing the same process, drawing gas onto resonant orbits due to the forcing frequencies of the bars. While it is not possible to distinguish between gas inflows and outflows in the plane of the discs using our bisymmetric radial bar-driven flow model, given the enhanced central molecular gas mass surface densities of our ``barred + radial flow'' subset galaxies, the preponderance of resonant inner orbits in these objects and our comparisons to the literature in this discussion, we are inclined to infer that we are observing radial bar-driven inflows as opposed to outflows.

\begin{figure*}
    \centering
    \includegraphics[width=0.47\linewidth]{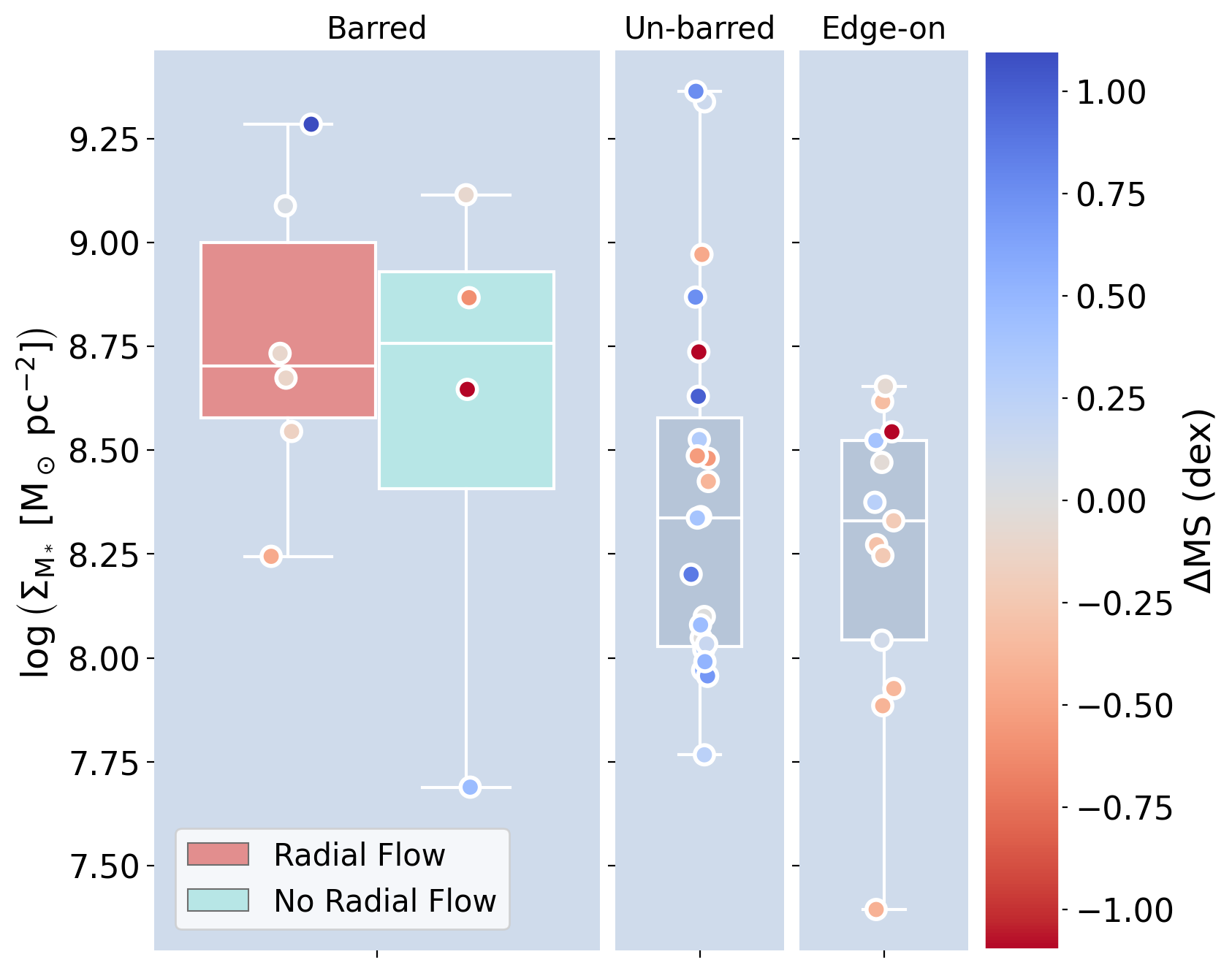}\hspace{10pt}
    \centering   
    \includegraphics[width=0.4\linewidth]{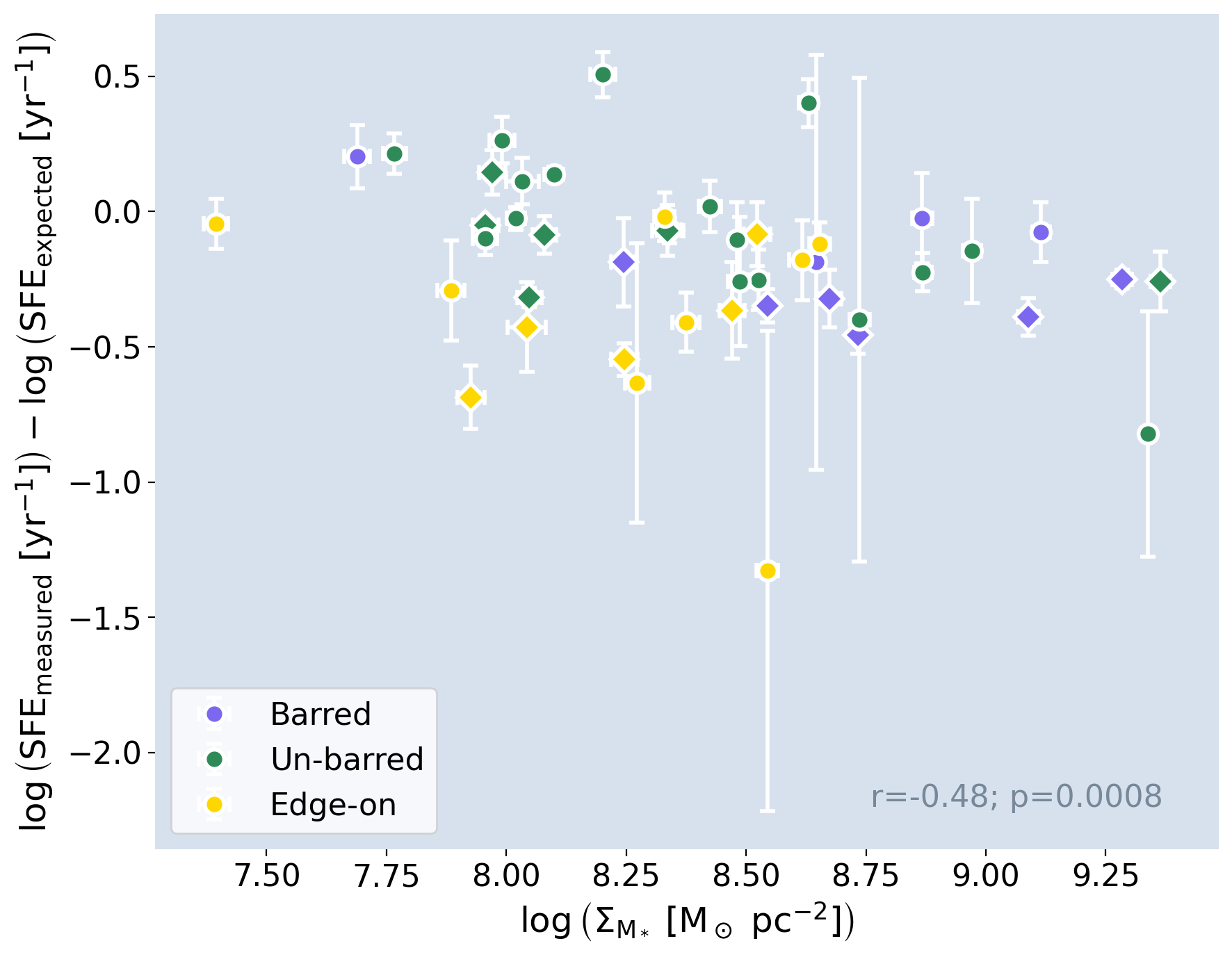}
    \caption{Distribution of the stellar mass surface densities ($\rm \Sigma _{M_*}$) in the central $\rm 1~kpc$ radius of of each of the morphological sub-samples. \textbf{Left}: the sample is divided into the three morphological sub-samples as in Figures~\ref{fig: vel_ratio}-\ref{fig: gas_density} and is split within the barred panel into radial bar-driven flow and no radial bar-driven flow subsets. The median position, inner quartiles and span of each distribution are shown by a box plot in each panel and the individual galaxies are shown as markers. The markers are colour-coded by their SFR offsets from the star-forming main sequence ($\rm \Delta MS$). \textbf{Right}: relationship between $\rm \log ({SFE}_{measured}) - \log ({SFE}_{expected})$ and central stellar mass surface density. Each marker is colour-coded by morphological sub-sample. Objects in the radial bar-driven flow subset are illustrated with a diamond-shaped marker. The Spearman correlation coefficient $\rm r=-0.48$ and the p-value $\rm <<0.05$, suggesting our $\rm \log ({SFE}_{measured}) - \log ({SFE}_{expected})$ quantities and the central stellar mass surface densities are negatively correlated.}
    \label{fig: sfe_stars}
\end{figure*}

\subsubsection{Un-barred galaxies}
\label{subsec: no bar gas flow}

While the focus of this paper is the molecular gas kinematics of galaxies hosting an optical bar, we also find some compelling results for our un-barred sub-sample. Only $0.30 \pm 0.10$ of un-barred galaxies are best modelled with radial bar-driven flows and the majority only require a 1-component surface brightness profile to describe the spatial distribution of their molecular gas ($0.70 \pm 0.10$; see Section~\ref{subsec: bars and gas}). When comparing these findings with those from the barred sub-sample, we can infer that these objects are kinematically distinct. Moreover, the results from Figure~\ref{fig: cons} further support this inference by illustrating the wide range of concentration parameters in the un-barred sub-sample and the higher median of the distribution. Once again we can defer to \citet{yu22} for a potential explanation for the wide range of concentration parameters that we measure. The authors find a correlation between the strength of non-axisymmetric structures and central molecular gas concentration, where axi-asymmetric structures can be due to a bar or the presence of spiral arms. They suggest that instead of there being a distinct difference between the gas dynamics in barred and un-barred galaxies, that the dynamics are a function of asymmetry, with a unified mechanism. This is further supported by \citet{geron21, geron23}, who extend this idea to different bar types, suggesting that there is a continuum of bar types, more specifically that there is no bifurcation between weak and strong bar in terms of the effects on their host galaxies. Our un-barred sub-sample likely contains galaxies with a range of structural asymmetries, corresponding to a range of pattern strengths with which to drive gas inwards. \citet{geron21, geron23} and \citet{walmsley22} also note that weak bars are under-detected in GZ2, with the barred sub-sample preferring strong bars. While our barred sub-sample is dominated by strong bars, the un-barred sub-sample is likely to contain a fraction of weakly barred galaxies. It is also possible that the un-barred sub-sample (as well as the other morphological sub-samples) contains galaxies under-going minor mergers or interactions, which can also act to centralise the molecular gas distributions \citep[e.g.][]{eliche-moral11}.

\subsection{Do radial bar-driven flows affect the SFEs of barred galaxies?}
\label{subsec: SFE and gas flow}

In Section~\ref{subsec: SFE}, we investigate the second goal of this paper, which is to look for correlations between non-circular gas kinematics and the SFEs of galaxies in our barred sub-sample. We define the quantity $\rm \log({SFE}_{measured}) - \log ({SFE}_{expected})$ (see Section \ref{subsec: SFE} for a detailed definition) to measure the enhancement/suppression of a galaxy's global SFE relative to the offset of SFE expected given its location on the star-forming main sequence (and the median global SFE of ALMaQUEST objects in the un-barred sub-sample within the scatter of the main sequence). In Figure~\ref{fig: del_sfe}, we illustrate the distribution of this quantity for each of our morphological sub-samples and radial flow subsets. Notably, we see an indication that the SFEs of the ``barred + radial flow'' subset galaxies are suppressed relative to the those of the un-barred sub-sample and ``barred + no radial flow'' subset galaxies. This result could imply that our ``barred + radial flow'' subset galaxies have suppressed total SFEs relative to those expected from their positions with respect to the star-forming main sequence. Echoing the distributions in Figure~\ref{fig: cons}, we also see a broader range of SFE offsets in the un-barred sub-sample, which could again be explained by the likelihood that this sub-sample is composed of a variety of galaxy morphologies, as argued in Section~\ref{subsec: no bar gas flow}. In order to investigate Figure~\ref{fig: del_sfe} further, in Figure~\ref{fig: sfe_disc} we plot the distributions of $\rm \log({SFE}_{H\upalpha, measured}) - \log ({SFE}_{expected})$ calculated in the discs of each of our morphological sub-sample galaxies, derived from MaNGA optical IFS (see Section~\ref{subsec: manga}). We find a similar distribution to that of Figure~\ref{fig: del_sfe} in terms of the relative enhancement/suppression between the morphological sub-samples and radial flow subsets. Again, our interpretation of this result is limited by the size of the samples, but finding a comparable result by an independent method does add credence to our interpretation of SFE suppression of the ``barred + radial flow'' subset galaxies. Extending this interpretation with Figure~\ref{fig: sfe_disc}, the suppression of SFE within the disc regions of galaxies in the ``barred + radial flow'' subset could be seen, in the context of the ``Compaction Scenario'' \citep{tacchella16}, as evidence of the disc being quenched by the inward radial flow of molecular gas. However, larger-scale analysis dividing barred galaxies into subsets depending on radial bar-driven flow detection is required to substantiate this argument.

\par In the literature, there is a general link between a galaxy being quenched and it having a higher stellar mass surface density ($\rm \Sigma _{M_*}$) in its central $\rm 1~kpc$ radius \citep[e.g.][]{fang13, woo15, davis22}. We assess this connection in Figure~\ref{fig: sfe_stars} and find that our barred sub-sample has a higher median central $\rm \Sigma _{M_*}$ than those of the other morphological sub-samples. Furthermore, all of the galaxies in the ``barred + radial flow'' subset have central $\rm \Sigma _{M_*}$ larger than $\rm \approx 8.25~M_\odot~pc^{-2}$, whereas the un-barred and edge-on sub-samples, as well as the ``barred + no radial flow'' subset, contain galaxies that have stellar mass surface densities below that threshold. Once again, this suggests a similarity within the ``barred + radial flow'' subset. It also is consistent with the literature, given that our most SFE-suppressed subset also has consistently high central $\rm \Sigma _{M_*}$, suggesting that growth of the central stellar mass surface density may be a signal of galactic quenching. 

\par It is widely theorised that molecular gas is depleted or expelled after large-scale starburst activity and exhausted in bulge-dominated early-type galaxies \citep[e.g.][]{gavazzi15, zolotov15, tacchella16, spinoso17}. However, in \citet{davis14, davis16} and \citet{french15}, this scenario is contradicted, the authors finding higher than expected gas fractions ($\rm \log M_{H2} / M_*$) in early-type galaxies. Furthermore, \citet{lin22} report in their analysis of green valley galaxies in ALMaQUEST that quenching is driven by the combination of reduced $\rm \log M_{H2} / M_*$ and suppressed SFE, challenging the picture that star formation is quenched purely by gas depletion. In Figure~\ref{fig: gas_dists}, we look at the distribution of molecular gas fractions in a 1~kpc central aperture of each of the morphological sub-samples. The ``barred + radial flow'' subset galaxies, despite having suppressed SFEs, do not appear to have depleted central molecular gas reservoirs, with the median central molecular gas fraction for the ``barred + radial flow'' subset galaxies roughly the same as that of the un-barred sub-sample galaxies (and enhanced compared to that of the ``barred + no radial flow'' subset galaxies). We conduct the same analysis using total molecular gas fractions and find a similar result. Moreover, in Figure~\ref{fig: gas_density} we measure a higher median central molecular gas mass surface density in the ``barred + radial flow'' subset, consistent with Figure~\ref{fig: cons} (i.e. in agreement with their centralised gas distributions). Using the 2-sided KS statistic, we find that the central molecular gas mass densities of the ``barred + radial flow'' subset galaxies are potentially drawn from a different distribution than those of the un-barred sub-sample (i.e. p-value < 0.05). Visually, it appears that the ``barred + radial flow'' subset galaxies are drawn from the higher-density end of the un-barred sub-sample's $\rm \Sigma _{H2}$ distribution and the ``barred + no radial flow'' subset from the lower-density end. We can infer from the higher median central molecular gas mass surface densities of the ``barred + radial flow'' subset galaxies that molecular gas depletion is not the main driver of SFR suppression of our ``barred + radial flow'' subset. This result is also supported by Figure~\ref{fig: sfe_disc}, with which we tentatively argue that the suppressed median disc SFEs in the ``barred + radial flow'' subset galaxies could be caused by inward radial bar-driven flows drawing gas away from their discs and effectively quenching star formation. Radial bar-driven flows of molecular gas appear to centralise the gas reservoirs of these objects, but do not fuel enhancements in SFR (i.e. neither globally or in the discs; we conduct a similar analysis for the central 1~kpc in Appendix~\ref{appendix: sfe}). In our un-barred sub-sample, however, the galaxies with similarly high central molecular gas mass surface densities have significantly higher $\Delta$MS than those in the ``barred + radial flow'' subset. This supports the argument that the presence of a bar is responsible for the SFE suppression instead of central gas depletion in these objects.

\par In order to explain the apparent relative suppression of SFEs in our ``barred + radial flow'' subset galaxies, we may require a more dynamical prescription of quenching. Dynamically-driven shear and the stability of the central gas reservoirs against collapse contribute to a morphological picture of quenching \citep[``morphological quenching'';][]{martig09}. For example, \citet{davis14} and \citet{gensior20} discuss the role of gas dynamics on SFE suppression in early-type galaxies (ETGs), in which the majority of the molecular gas reservoirs are concentrated in the rising portions of the galaxies' rotation curves, where shear is generally higher. Although our barred sub-sample galaxies are clearly morphologically distinct from ETGs, the ``barred + radial flow'' subset does contain consistently centralised molecular gas reservoirs, implying that the gas is restricted to the high-shear regions of the host galaxies. Furthermore, the presence of radial flows will also contribute to shear, particularly in the central areas where most of the gas is concentrated. The presence of strong shear could also increase the velocity dispersion of cold gas clouds by actively pulling them apart, increasing the Toomre Q parameter \citep{toomre64}. This is corroborated by the studies of M51 by \citet{meidt13b, meidt13a}, who detect a clear anti-correlation between strong gas flows and star formation (traced by H$\upalpha$ and 24$\upmu$m emission). This implies a direct connection between the star-forming ability of giant molecular clouds (GMCs) and their dynamical environment, suggesting that GMCs in the vicinity of radial streaming motions are significantly stabilised against gravitational collapse. These findings contribute to a scenario where our ``barred + radial flow'' galaxies have driven gas inwards (via the radial flows induced by bars) towards central molecular gas reservoirs that are largely gravitationally stable. This is in comparison to our ``barred + no radial flow'' subset galaxies, which are less centrally concentrated and obviously do not have the contribution to shear provided by radial flows. As discussed in Section~\ref{subsec: bar gas flow}, these objects are potentially at different dynamical stages of their evolution (e.g. prior to the radial inflow of gas or after a central starburst once the central molecular gas reservoir gravitationally destabilises). As depicted in Figures~\ref{fig: del_sfe} and \ref{fig: cons}, our most concentrated objects are largely quenched, with SF restricted to the central regions and maybe also subject to dynamical suppression.

\par While there are multiple studies in the literature that find a connection between galaxies hosting bulges and SFE suppression \citep[e.g.][]{saintonge12, martig13, davis14}, the influence of bars on the star-forming potential of galaxies is less consistent. \citet{wang12} summarise the variety of results in this field, finding that the galaxies in their sample with strong bars either have enhanced central star-formation rates or star formation that is suppressed. Resolved studies find that the SFE can vary along a bar and between galaxies; e.g. \citet{maeda23} finds star formation generally suppressed along bars but enhanced at the bar ends, while \citet{diaz-garc21} find no evidence of SFE variation between the bar and bar ends but significant variation between galaxies. These differences may be explained to some extent by bar/bar-end definitions in the case of resolved studies, but the variety of results may also indicate that galaxies classed as barred in these studies contain a mixture of objects dominated by different mechanisms or at different stages of their evolution. Our identification of radial bar-driven flows may provide a method to sub-divide barred galaxies into subsets with similar global properties based on the large-scale dynamics of their molecular gas. We suggest that this sub-division of optically-barred galaxies potentially captures the portion of objects with suppressed SFEs in the literature. Without making the distinction, many studies find little difference in the star-formation efficiency of barred galaxies when comparing them to their un-barred counterparts \citep[e.g.][]{saintonge12}. To assess this result further, a larger sample of galaxies with CO data of comparable resolution and sensitivity to the data of ALMaQUEST would be required. 


\section{Summary \& conclusions}
\label{sec: conclusions}

In this work, we investigate the statistical relationship between the presence of an optical bar and radial molecular gas flows and, furthermore, whether that connection influences the star-formation activities of the host galaxies. We use CO(1-0) maps of 46 galaxies from the ALMaQUEST survey to create 3D kinematic models (using the \textsc{KinMS} package), from which we infer whether or not there is evidence of radial gas motions. In tandem with our dynamical analyses of the molecular gas in this sample, we also create morphological classes (i.e. barred, un-barred and edge-on galaxies) based on optical classifications from Galaxy Zoo and HyperLEDA. By combining our morphological sub-samples with our radial bar-driven flow classifications, we find the following key results:

\begin{itemize}

    \item We find a bar fraction of $0.30 \pm 0.08$ for the ALMaQUEST sample using our classification procedure outlined in Figure~\ref{fig: bar_class}.

    \item In Figure~\ref{fig: bar_rad_frac}, we show that $0.60 \pm 0.15$ of the barred sub-sample galaxies are classed as having radial bar-driven flows, compared to $0.30 \pm 0.10$ and $0.38 \pm 0.13$ of the un-barred and edge-on sub-samples, respectively. This suggests that gas flows resembling radial bar-driven flows are more common in galaxies hosting axi-asymmetric features like bars. 

    \item $0.90 \pm 0.09$ of our barred sub-sample galaxies have a multi-component surface brightness profile as the best-fit model to their molecular gas distribution (see Figure~\ref{fig: morph_frac}). This is likely indicative of resonant ring structures driven by bars. We also find a greater proportion of our barred sub-sample galaxies have a multi-component surface brightness profile compared to the barred sub-sample galaxies hosting radial bar-driven flows. This supports a scenario where gas inflows are shorter lived than the timescale over which galactic bars are stable.

    \item Comparing the radial and transverse velocity components ($\overline{V}_{\rm r}$ and $\overline{V}_{\rm t}$) of the radial bar-driven flows used in our models (defined in Section~\ref{subsec: models}), in Figure~\ref{fig: vel_ratio} we find that our barred sub-sample galaxies have $\overline{V}_{\rm r}\ /\ \overline{V}_{\rm t} \lesssim 1$, but there are much wider spreads of $\overline{V}_{\rm r}\ /\ \overline{V}_{\rm t}$ ratios in our other morphological sub-samples. This not only implies a dynamical similarity of the molecular gas kinematics in our barred objects, but also suggests that the gas is spiraling inwards, as opposed to just moving radially, in these objects.

    \item The objects in our ``barred + radial flow'' subset appear to have suppressed SFEs compared to those of the ``barred + no radial flow'' subset and in the un-barred sub-sample (both globally using GSWLC-X2 and WISE values and in their discs using MaNGA IFS; see Figures~\ref{fig: del_sfe} and \ref{fig: sfe_disc}). The ability to make this distinction, through molecular gas kinematics, goes some way in explaining the tension between various results in the literature regarding SFE-enhancements/suppression in barred galaxies.

    \item Galaxies in our ``barred + radial flow'' subset have consistently concentrated molecular gas distributions compared to those of the other subsets, which all have a wider spread of concentration parameters (see Figure~\ref{fig: cons}). Again, we interpret this as an indication that the galaxies in the ``barred + radial flow'' subset are undergoing a similar kinematic process, whereby gas is driven inwards.

    \item In Figure~\ref{fig: gas_dists}, we show that the objects in the ``barred + radial flow'' subset are not depleted of molecular gas relative to the other subsets when comparing their respective central molecular gas mass fractions. These galaxies also has a higher median molecular gas mass surface density in their inner 1~kpc than the ``barred + no radial flow'' subset (see Figure~\ref{fig: gas_density}), suggesting that radial bar-driven inflows may act to centralise gas in these galaxies. Exhaustion of the central molecular gas reservoirs is, therefore, not the reason for the SFR suppression observed in this subset. We speculate that dynamical effects, such as shear, driven by the radial motions of gas may act to quench the centralised gas reservoirs. 

\end{itemize}

\par In conclusion, we find that barred galaxies are more likely to host radial molecular gas flows, and that the presence of those flows can also alter their large-scale properties, notably their star-forming efficiencies and the central compaction of their molecular gas reservoirs. Despite the molecular gas being more centrally concentrated in barred galaxies with detected radial bar-driven inflows, they do not power central SFR enhancements (at least not while the flows are on-going). Conversely, we find these galaxies appear to have suppressed SFEs both globally and in their discs. From this, we can begin to infer that gas is being drawn inwards, quenching the discs of these galaxies, and increasing their central molecular gas mass surface densities. However, the dynamics of the flows themselves may act to stabilise the molecular gas against fragmentation and collapse (e.g. through effects like shear), suppressing the SFEs of the galaxies' molecular gas reservoirs.

\par To further this work, a larger survey of galaxies with high spatial and spectral resolution CO mapping and with high sensitivity is required, as to date these surveys have been limited in size. The resolutions and sensitivity of the data are critical for meaningful kinematic modelling, which we propose is a vital tool for understanding the baryon cycle and its impact on the evolution of barred galaxies. 


\section*{Acknowledgements}
\label{sec: acknowledgements}

LMH and AS acknowledge support from the Royal Society. TAD acknowledges support from the UK Science and Technology Facilities Council through grants ST/S00033X/1 and ST/W000830/1. This paper makes use ALMA data from project S/JAO.ALMA \#2011.0.01234.S. ALMA is a partnership of ESO (representing its member states), NSF (USA) and NINS (Japan), together with NRC (Canada), MOST and ASIAA (Taiwan), and KASI (Republic of Korea), in cooperation with the Republic of Chile. The Joint ALMA Observatory is operated by ESO, AUI/NRAO and NAOJ.

LL acknowledges support from the National Science and Technology Council (NSTC) through grants MOST 108-2628-M-001-001-MY3 and NSTC 111-2112-M-001-044-. HAP acknowledges support by the National Science and Technology Council of Taiwan under grant 110-2112-M-032-020-MY3.

This project makes use of the MaNGA-Pipe3D data products. We thank the Instituto de Astronomía - Universidad Nacional Autonoma de Mexico (IA-UNAM) MaNGA team for creating this catalogue, and the Conacyt Project CB-285080 for supporting them.

\section*{Data availability}

All ALMaQUEST data products used in this work are available in the ALMA science archive (\url{https://almascience.eso.org/asax/}). MaNGA IFS and stellar mass maps are also publicly available in the SDSS Science Archive Server (SAS; \url{https://dr17.sdss.org/sas/}). GALEX and WISE data can be acquired from their respective online databases (i.e. \url{https://salims.pages.iu.edu/gswlc/} and \url{http://vizier.cds.unistra.fr/} respectively). The NSA data used in this paper can be accessed via the SDSS DR17 website (\url{https://www.sdss4.org/dr17/manga/manga-data/catalogs/#DRPALLFile}).



\bibliography{refs.bib}
\bibliographystyle{mnras}


\appendix

\section{Central SFEs derived from MaNGA}
\label{appendix: sfe}

\begin{figure}
    \centering
    \includegraphics[width=\linewidth]{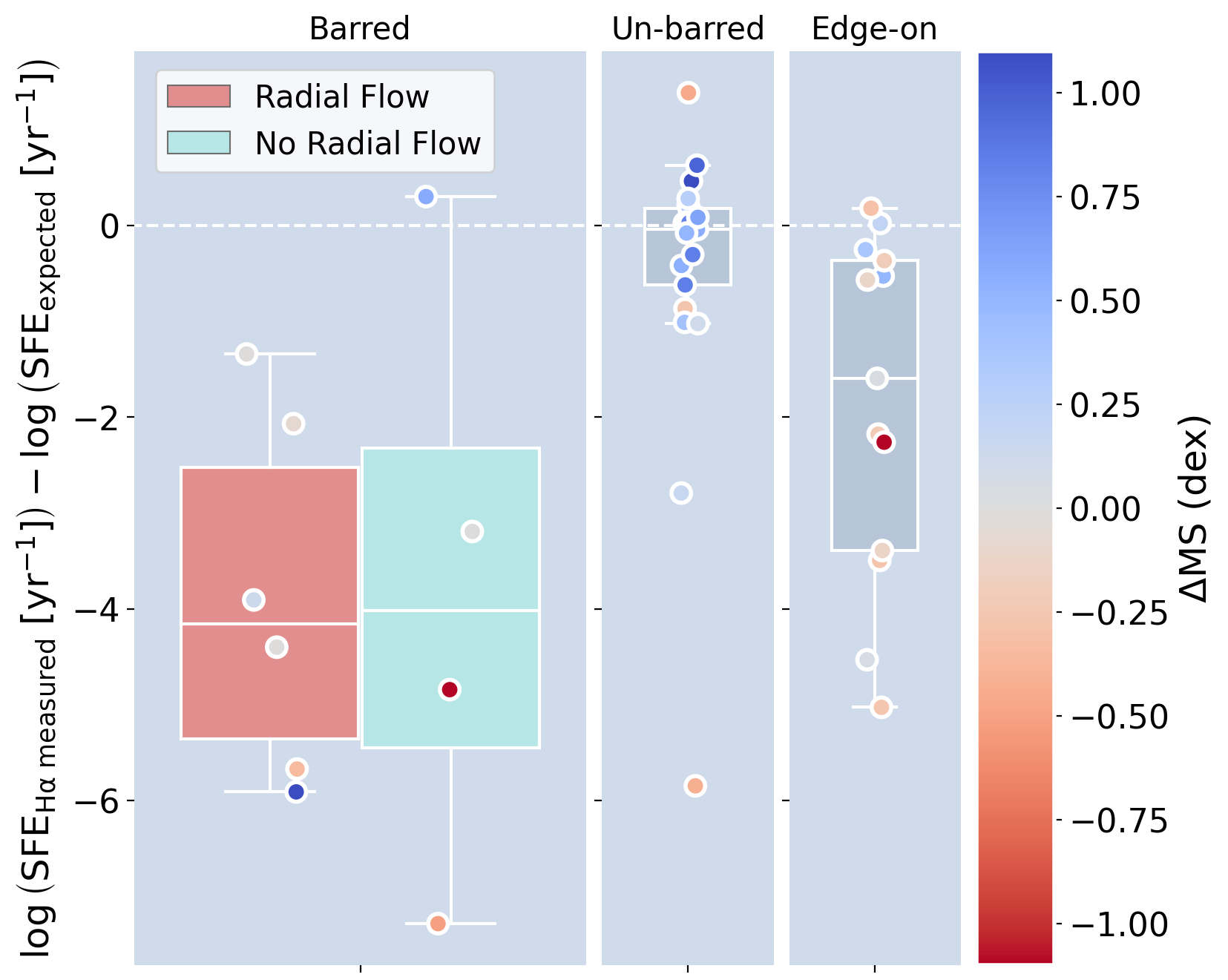}
    \caption{Distributions of the $\rm {SFE}_{H\upalpha, measured} - {SFE}_{expected}$ quantities in the central regions of galaxies in each of our morphological sub-samples using H$\upalpha$ emission from MaNGA. For each object, we sum SFR and $\rm M_{H2}$ spaxels within the central 1~kpc regions to calculate the median central SFE.}
    \label{fig: med_sfe}
\end{figure}

In Figure~\ref{fig: med_sfe}, we illustrate differences of $\rm {SFE}_{H\upalpha,measured} - {SFE}_{expected}$ of the galaxies in our morphological sub-samples and radial flow subsets in their central 1~kpc regions. As previously stated, this analysis is limited by the number of spaxels masked in the central region of each object. We, therefore, take $\rm {SFE}_{H\upalpha,measured}$ as the median SFE spaxel value in this region, using the MaNGA SFR maps and ALMaQUEST $\rm M_{H2}$ maps re-binned to the MaNGA spaxel grid.

\section{SNR degradation}
\label{appendix: snr}

We test for variations of the S/N in the ALMaQUEST datacubes by degrading all cubes with S/N > 30 to below this threshold. S/N$\approx$30 is the median S/N of objects in the un-barred sub-sample. We degrade these cubes by creating 2D images of randomly generated values drawn from a Gaussian distribution of width equal to the standard deviation of the original cube (in line-free regions). We smooth these 2D noise images with the synthesised beam and add them to each channel of the original datacubes (each 2D noise image is created independently for each channel). The new S/N of the degraded cube is then calculated and if it is still > 30, we create 2D noise images with integer multiples of the original cube standard deviation and repeat the process, with increasingly higher multiples of the noise, until its S/N < 30. 

\begin{figure}
    \centering
    \includegraphics[width=\linewidth]{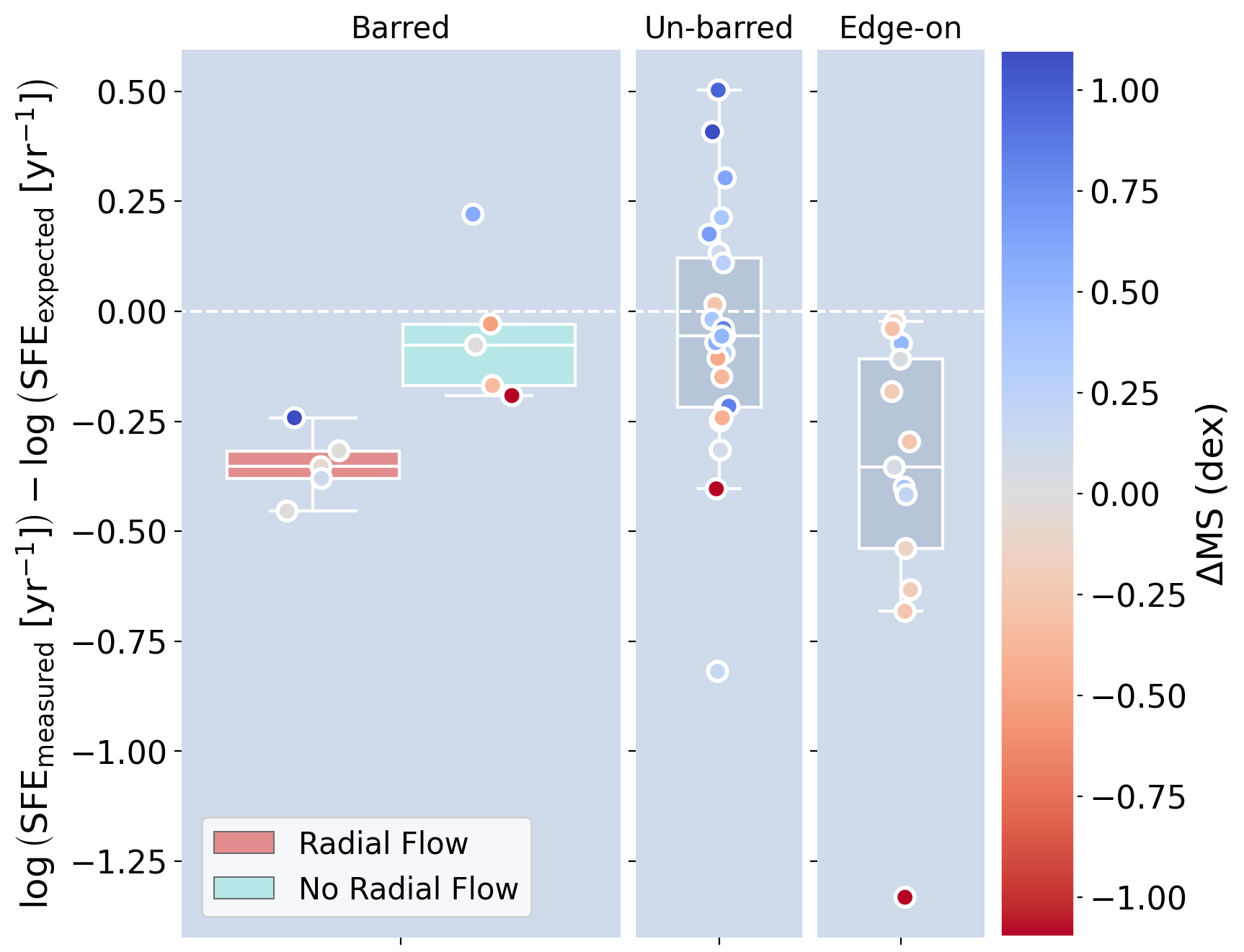}
    \caption{Figure~\ref{fig: del_sfe} re-plotted using our S/N-degraded datacubes and \textsc{KinMS} models.}
    \label{fig: deg_sfe}
\end{figure}

\begin{figure*}
    \centering
    \includegraphics[width=0.45\linewidth]{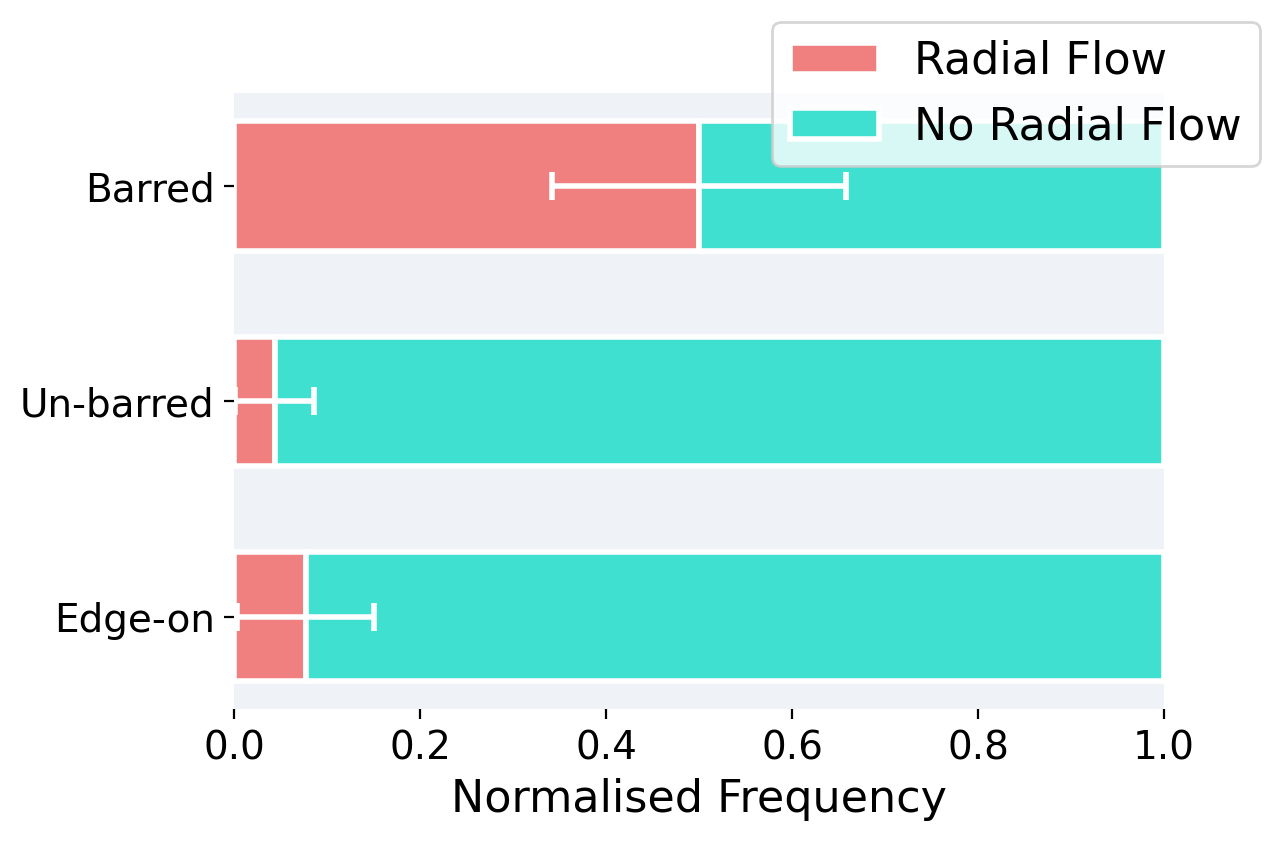}
    \centering
    \hspace{10pt}
    \includegraphics[width=0.45\linewidth]{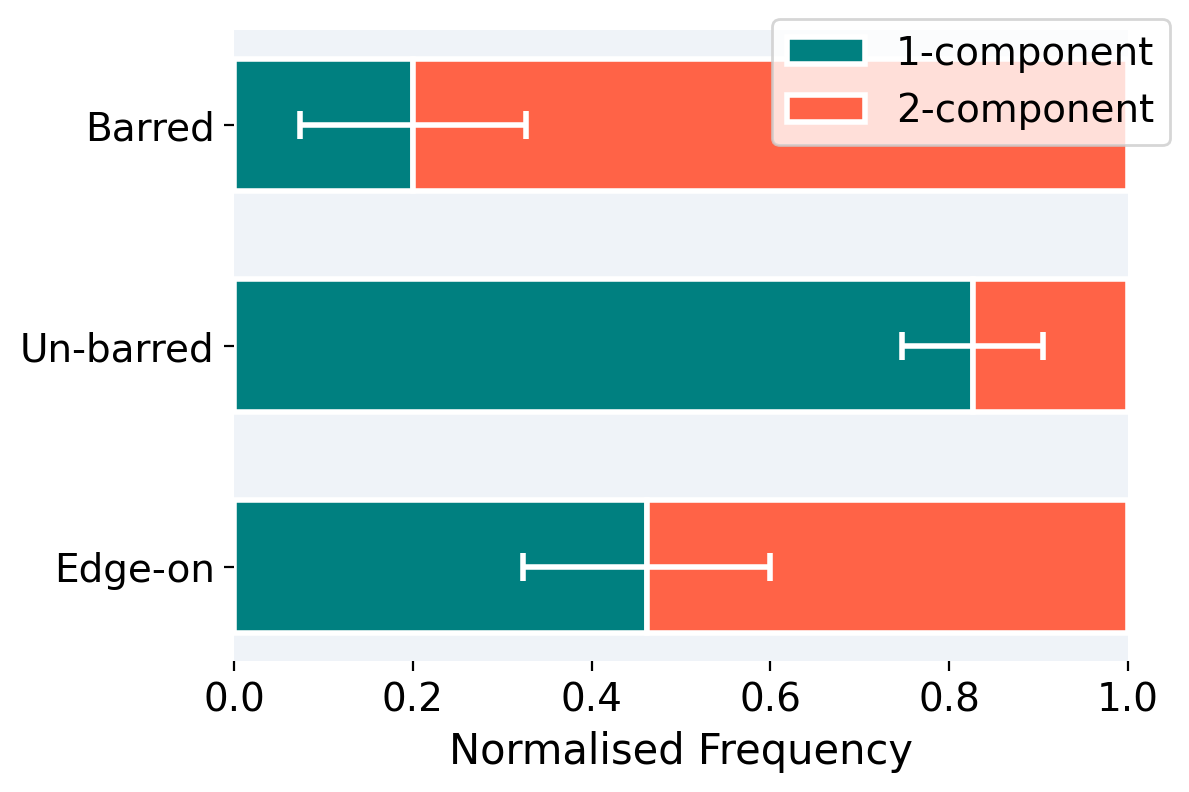}
    \caption{Figures~\ref{fig: bar_rad_frac} (left) and \ref{fig: morph_frac} (right) re-plotted using KinMS models created using our S/N-degraded datacubes.}
    \label{fig: deg_fracs}
\end{figure*}

\par Our degraded datacubes are re-modelled using \textsc{KinMS}, replicating the process used for the original cubes. This produces a new distribution of best-fit surface brightness profiles and a new radial bar-driven flow subset. The fraction of galaxies of each morphological sub-sample with radial bar-driven flows is presented in Figure~\ref{fig: deg_fracs} alongside the fraction with 2-component surface brightness models. The fraction of barred galaxies with radial bar-driven flows is $0.50 \pm 0.16$, compared to $0.04 \pm 0.04$ and $0.08 \pm 0.07$ for the un-barred and edge-on sub-samples, respectively. The fraction of barred galaxies with 2-component models is $0.80 \pm 0.13$, compared to $0.17 \pm 0.08$ and $0.54 \pm 0.14$ for the un-barred and edge-on sub-samples, respectively. 

\par The S/N of the datacubes, therefore, do affect the detection of radial bar-driven flows and the best-fit model. However, they do not affect our conclusions. Figure~\ref{fig: deg_sfe} illustrates this further by showing the suppression of SFE in the ``barred + radial flow'' subset relative to that of the ``barred + no radial flow'' subset, replicating our original interpretation. We infer, therefore, that our results and conclusions are robust against S/N variations in the ALMaQUEST data.

\bsp	
\label{lastpage}
\end{document}